%% file: paper.tex
\documentclass[sigconf,natbib=false]{acmart}

\AtBeginDocument{%
  }

\RequirePackage[
  datamodel=acmdatamodel,
  style=acmnumeric,
  ]{biblatex}

\usepackage{tikz}
\usepackage{amsmath}

\usepackage{hyperref}
\usepackage{graphicx}
\graphicspath{ {./images/} }   

\newcommand{\systemName}[1]{UsernameCrazy}
\newcommand{\frameworkName}[1]{SQUAD} 
\newcommand{\ignore}[1]{}

\usepackage[skip=0pt]{caption}

\usepackage[colorinlistoftodos]{todonotes}

\usepackage{tablefootnote}

\usepackage{multirow}

\usepackage{algorithm}
\usepackage[noend]{algpseudocode}

\usepackage{epsfig}

\usepackage{tcolorbox} 

\usepackage{soul}
\usepackage{framed}

\usepackage[final]{changes}

\begin{document}

\title{Username Squatting on Online Social Networks: A Study on X}

\definecolor{brinkpink}{rgb}{0.98, 0.38, 0.5}




\author{Anastasios Lepipas}
\affiliation{%
 \department{Department of Computing}
 \country{Imperial College London, UK}
    }
\email{a.lepipas20@imperial.ac.uk}

\author{Anastasia Borovykh}
\affiliation{%
  \department{Department of Computing}
  \country{Imperial College London, UK}
    }
\email{a.borovykh@imperial.ac.uk}

\author{Soteris Demetriou}
\affiliation{%
    \department{Department of Computing}
  \country{Imperial College London, UK}
    }
\email{s.demetriou@imperial.ac.uk}

\input{sections/abstract}

\ccsdesc[500]{Information systems~Social networks}
\ccsdesc[500]{Security and privacy}

\keywords{username squatting, social networks, impersonation, typo-mentions}


\maketitle

\input{sections/1_introduction}
\input{sections/2_problem_statement}
\input{sections/3_measurement_study}
\input{sections/4_results}
\input{sections/5_design}
\input{sections/6_evaluation}

\input{sections/7_applying_squad}
\input{sections/8_background_work}
\input{sections/9_conclusion}



\printbibliography


\input{sections/appendix}

\end{document}

%% file: sections/abstract.tex
\begin{abstract}
Adversaries have been targeting unique identifiers to launch typo-squatting, mobile app squatting and even voice squatting attacks. Anecdotal evidence suggest that online social networks (OSNs) are also plagued with accounts that use similar usernames. This can be confusing to users but can also be exploited by adversaries. However, to date no study characterizes this problem on OSNs.
In this work, we define the \textit{username squatting} problem and design the first multi-faceted measurement study to characterize it on \replaced[]{\emph{X}}{Twitter}. We develop a username generation tool (\systemName{}) to help us analyze hundreds of thousands of username variants derived from celebrity accounts. Our study reveals that thousands of squatted usernames have been suspended by \replaced[]{\emph{X}}{Twitter}, while tens of thousands that still exist on the network are likely bots. Out of these, a large number share similar profile pictures and profile names to the original account signalling impersonation attempts. We found that squatted accounts are being mentioned by mistake in \textit{tweets} hundreds of thousands of times and are even being prioritized in searches by the network's search recommendation algorithm exacerbating the negative impact squatted accounts can have in OSNs.
We use our insights and take the first step to address this issue by designing a framework (\frameworkName{}) that combines \systemName{} with a new classifier to efficiently detect suspicious squatted accounts. Our evaluation of \frameworkName{}'s prototype implementation shows that it can achieve $\textbf{94\%}$ F1-score when trained on a small dataset.
\end{abstract}

\keywords{username squatting, social networks, typo mentions, impersonation}

%% file: sections/1_introduction.tex
\section{Introduction}
\label{sec:introduction}
Users in online social networks (OSNs) are distinguished through unique identifiers or \textit{usernames}. These identifiers are used for searching for and interacting with other users. Anecdotal evidence suggest that similar usernames might lead to confusion on the network. To make things worse, adversaries---who have been known to opportunistically hijack popular OSN accounts~\cite{Profile_Name_Reuse}---can deliberately weaponize this feature to impersonate popular or influential users to further their malicious campaigns and advertise content, products or services, spread propaganda, hate content or fake news, or harm influential individuals~\cite{Low_credibility_by_social_bots,Characterizing_Social_Bots}.

Adversaries have targeted unique identifiers in the past using a range of squatting techniques with great success. For instance, traditional typo-squatting attacks target users who \ignore{incorrectly type}mistype a URL~\cite{URLCrazy}, mobile-app squatting attacks create mobile apps with names similar to popular apps to increase their installation counts and otherwise fool mobile users~\cite{Mobile_App_Squatting}, and voice squatting attacks target phonetic similarities between the invocation names of voice assistant applications to hijack their communication with their users~\cite{Skill_Squatting_Alexa}. 

However, there is no systematic measurement of similar phenomena on OSNs. Prior works focused on analyzing and detecting fake accounts, content and bots on OSNs. These fail to study the extent of squatting on OSNs and mostly rely on expensive graph traversals to identify Sybil nodes~\cite{Doppelganger_Bot_Attack} resulting in non practical detection methodologies. The rise of disinformation and fake accounts on OSNs already forced regulators to employ stricter policies---such as the \textit{Code of Practice on Disinformation}~\cite{code_practice_disinformation_2022} backed by the European Union's \textit{Digital Services Act}~\cite{digital_services_act}---to hold OSNs accountable for disinformation and fake accounts on their platforms. Therefore, it is timely and paramount to better understand squatting on OSNs and study techniques for tackling this problem.

Our work is the first which studies \textit{username squatting} on OSNs. We define and characterize \textit{username squatting}, and develop an efficient and effective methodology for identifying squatted accounts on OSNs. To achieve our first goal, we design a systematic measurement study for one of the most popular networks~\cite{Twitter_Statistics}, \emph{X} (previously known as Twitter~\cite{x_rebrand}). Our study is structured around four measurement questions: (\textbf{MQ$\textbf{1}$}) Is \textit{username squatting} a prevalent problem on \replaced[]{\emph{X}}{OSNs}? If it is, then (\textbf{MQ$\textbf{2}$}) how does it contribute to online confusion; (\textbf{MQ$\textbf{3}$}) what are the username characteristics of the most problematic username variants, and (\textbf{MQ$\textbf{4}$}) what is the specific behavior of the squatted accounts (i.e. what is the \emph{purpose} of the owner of such accounts)? 

To aid our analysis, we developed a tool, \systemName{}, inspired by typo- and mobile app- squatting methodologies~\cite{URLCrazy, Mobile_App_Squatting} but tailored to the characteristics of OSN accounts, which uses $10$ string generation strategies to efficiently construct more than $851,000$ username variants for the top $97$ \added[]{(in $2019$)} popular \replaced[]{\emph{X}}{Twitter} accounts~\cite{kaggle_dataset_twitter}, for which we collected more than $1.2$ million tweets in total. \added[]{$83$ out of $97$ are still in top $100$ most popular accounts in 2023~\cite{top_popupal_accounts_on_X_2023, top_popupal_accounts_on_X_2023_2}.} Our measurement study led to a number of important findings. We found that tens of thousands of potentially confusing variants exist on the network; a large number of username variants of popular accounts are \emph{mistakenly} mentioned (typo-mentioned) and that confusing variants appear high in the OSN's search recommendation list---and at times even \emph{before} the verified account (see \replaced[]{Section~\ref{sec:online_confusion}}{Figure~\ref{fig:suggested_user} in Appendix~\ref{appendix:more_examples}}). We found that variants with a small edit distance are more likely to be mentioned and suggested. Both phenomena lead to increased attention for these confusing accounts which malicious actors can consequently exploit. As proof of this, we observed that a subset of these variants are already suspended which suggests that \textit{username squatting} is indeed leveraged for malicious purposes. Moreover our three data collection snapshots across one year show that this is a growing problem as we found the number of suspended variants to increase over time. Analysing the squatted accounts, we show that thousands of squatted accounts which are currently \emph{active} on the network are likely bots. A significant amount is involved in selling products, political goals or spreading fake news. Thus, as expected, the confusion generated through similar usernames and profiles tends to be exploited with malicious intentions. 

Our \deleted[]{measurement }study further shows that a large amount of squatted username accounts are involved in \emph{previously unknown} impersonation attacks;\ignore{see for an example Figure \ref{fig:David_Guetta_Impersonation}.} To tackle \replaced[]{this}{such issues}, \replaced[]{\emph{X}}{Twitter} recently launched a premium service which allowed everyone paying a small fee ($\$8$ per month) to be granted a blue check on their account~\cite{8_per_month}. Blue checks signal account verification and this feature was quickly and heavily manipulated by a swath of fake accounts to spread misinformation~\cite{fake_misinformation} and even damage companies' brands and stock prices~\cite{fake_EliLilly, fake_Martin}. Companies reported to be withdrawing their ad campaigns from \replaced[]{\emph{X}}{Twitter} due to the uncertainty of the platform's directions and to be pausing their \replaced[]{\emph{X}}{Twitter} publishing plans for all their corporate accounts~\cite{twitterCost_EliLilly, Forbes_Ads, CNN_ads}. The \deleted[]{premium }feature was reverted and \replaced[]{\emph{X}}{Twitter} claimed that blue checks will be granted only after manual verification~\cite{twitterManualVerification}. The platform \deleted[]{has now }relaunched a modified version of their premium service which uses color--coded badges for different entities (organisations vs individuals etc.) along with an increased fee ($\$11$ per month) if registered from a mobile device~\cite{8_per_month}. However, anecdotal evidence suggest that color badges are not a foolproof method, allowing adversaries to add cloned badges next to their profile name~\cite{advers_color_badge1,advers_color_badge2}
Also, manual verification is cumbersome and costly which is further exacerbated by the fact that OSNs currently operate with restricted resources~\cite{Twitter_layoffs,Meta_layoffs}.

Our measurement study revealed that profiles sharing many similar features with the original account may try to exploit the created confusion and generate traction to their profile to promote products or services, or share potentially malicious links. This led us to our design question: (\textbf{DQ}) Can we design an effective and
efficient framework for detecting suspicious username squatting attempts? Toward this, we leverage our previous observations to design \textbf{\frameworkName{}}. \frameworkName{} combines user name generation and online account discovery with the classification of suspicious squatted accounts. By ensuring a low false positive rate, we envision our framework to be an effective tool in scanning the OSN, saving time in manual screening and resulting in the more rapid discovery of suspicious profiles. A prototype of \frameworkName{} trained on a small dataset reaches an F1-score of $94\%$ in detecting suspicious accounts. We further applied \frameworkName{} on popular and non-popular accounts and found that squatting targets popular users (55\% of accounts squatting popular users are classified as malicious). Upon analyzing their activity we reveal two trends: (a) posting insecure URLs and at a lesser but non-negligible extend (b) trying to amplify their follower base.   Our findings were responsibly disclosed (see Appendix~\ref{appendix:ethics} for details). \added[]{\frameworkName{} is available on GitHub~\cite{SQUAD_Github}).} \added[]{Examples of impersonators, confusing variants, fake tweets and suspicious links, and additional results are available on our project's website~\cite{SQUAD_Google_Site}}.

\vspace{3pt}\noindent\textbf{Contributions.} We summarize our main contributions below:

\vspace{3pt}\noindent$\bullet$ \textit{New techniques.} We propose \systemName{}, an efficient tool embodying new techniques for generating squatted versions of OSN usernames. 

\vspace{3pt}\noindent$\bullet$ \textit{Findings.} We performed a measurement study and found that username squatting is a prevalent and growing problem, it can contribute to online confusion, and is used for malicious activity.

\vspace{3pt}\noindent$\bullet$ \textit{Detection Framework.} We design, implement, evaluate and apply \frameworkName{}, a novel end-to-end framework that leverages username squatting as a strong signal to efficiently and effectively discover suspicious squatted accounts.

%% file: sections/2_problem_statement.tex
\section{Problem Statement}
\label{sec:problem}

\vspace{3pt}\noindent\textbf{What is username squatting?} OSNs allow their users to create accounts with ease. Users can select any available string for their username\footnote{There is no mention that username policies, in contrast with suspension policies~\cite{Country_Withheld_Content, Country_Settings}, are not the same globally.}, which can be used to uniquely identify them across the network. Other users can leverage those usernames to search, mention and interact with account owners. However, this poses a significant threat, as a malicious actor can easily select identifiers at account creation time which are confusingly similar to targeted entities or individuals for malicious purposes. Such techniques have been applied in the past to impersonate web domains (URL squatting~\cite{URLCrazy, Typo_Squatting}), mobile apps (mobile app squatting~\cite{Mobile_App_Squatting}) and even voice personal assistant skills (skill squatting~\cite{Skill_Squatting_Alexa, zhang2019dangerous}) with very high success rates. Similarly, a malicious actor could select a username which is similar with usernames of popular accounts in an attempt to impersonate the target account and/or spread confusion. We call this adversarial strategy \emph{username squatting}. We will refer to squatted usernames as account usernames that are confusingly similar to popular account usernames (such as the ones of verified accounts). Similarly, we will refer to accounts that use squatted usernames as squatted accounts.


\vspace{3pt}\noindent\textbf{Definitions.} We distinguish benign and suspicious accounts with the latter split into confusion and impersonation.

\vspace{3pt}\noindent\textit{Benign.} Benign accounts have an account name which is a username variant of an original account. However, these users either do not share a profile name, or their profile image is not similar to the original account, or they explicitly mention that they are a fan or parody \ignore{accounts}users which by \replaced[]{\emph{X}}{Twitter}'s policy is allowed~\cite{Twitter_policies}. 

\begin{figure}[!htb]
\centering
\includegraphics[width=0.9\columnwidth,height=\textheight,keepaspectratio]{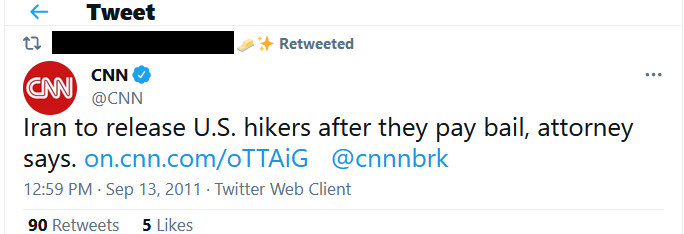}
\caption{Typo-mention: CNN mistakenly mentions \textit{@cnnnbrk}, a squatted user of the official \textit{@cnnbrk} profile.}
\label{fig:Tweet1}
\vspace{-1em}
\end{figure}

\vspace{3pt}\noindent\textit{Confusion.} We define \emph{confusion} accounts on OSNs to be squatted accounts which share a sufficient number of features with the original accounts and are suspicious (i.e., clearly not benign). Bot accounts are a subset of this class. Confusion can be exploited for malicious purposes (e.g., share links, fake news) and their impact can be amplified through features of OSNs (e.g., the search engine, likes, mentions). Figure~\ref{fig:Tweet1} illustrates a scenario where a verified account on \replaced[]{\emph{X}}{Twitter} (identified by the blue check on its name~\cite{Twitter_Badge}) of a prominent news network mistakenly mentions a fake confusion account ``cnnnbrk'', which is \textit{liked} and \textit{retweeted} by numerous users. 

\vspace{5pt}\noindent\textit{Impersonation.}
\label{subsec:problem_statement_impersonation}
We define accounts to be \emph{impersonation} attempts if these have the same features as the original accounts \emph{and} impersonate their \emph{behavior} either through the tweets or information contained in the bio (e.g., fake verification blue check). Impersonations are a subset of the broader confusion class. While we focus on discovering the more general confusion accounts, we do find a number of impersonations as well\deleted[]{ (see  Figures~\ref{fig:David_Guetta_Impersonation},~\ref{fig:malicious_links_profile},~\ref{fig:theellenshow_impersonation},~\ref{fig:shakira_impersonation} in Appendix~\ref{appendix:more_examples} for examples)}.

%% file: sections/3_measurement_study.tex
\section{Measurement Study Design}
\label{sec:measurement_study}

\subsection{Data Collection and Username Generation.}
\label{subsec:usernameCrazy}
\label{subsec:automated_username_variant}

\vspace{3pt}\textbf{Ethics in Data Collection.} We analyzed data available through \replaced[]{\emph{X}}{Twitter}. We did not extract any personal opinions or other viewpoints linked to individuals as this could potentially be sensitive. In detail, a) all our collected raw data (profile data, tweets with mentions etc.) were captured using official APIs and b) all the unnecessary data that APIs return are discarded immediately. We also b) never share any information with $3$rd parties and c) encrypt and keep in a local disk all the data after each experiment. \replaced[]{Besides, we}{We furthermore} follow recommendations in~\cite{williams2017towards} and mask the account names of individuals in the paper and do not include any tweets of individual accounts, including only public tweets from the top $97$ most popular users.

\vspace{5pt}\noindent\textbf{Data Collection Methods.}\label{subsec:data_collection}
We focus our analysis on \replaced[]{\emph{X}}{Twitter} due to its popularity. \replaced[]{\emph{X}}{Twitter} has more than \replaced[]{$353$}{$192$} million \added[]{monthly} active users in 2023~\cite{Twitter_Statistics}\ignore{, the fact that it allows for discovering suspended accounts and because there exist tools which allow us to look for bot-like behavior}. 
We select the $100$ most popular accounts of \replaced[]{\emph{X}}{Twitter}~\cite{kaggle_dataset_twitter}. We call this dataset the ‘\textit{Initial Seed}’. Out of those accounts, at the data collection time ‘@realdonaldtrump' had been suspended while ‘@aamir\_khan' and ‘@arianagrande' had been deleted or deactivated. 
We exclude these accounts from our dataset and all measurements. This is because \replaced[]{\emph{X}}{Twitter} API does not return information about suspended, deactivated and deleted accounts. This would pose a threat to validity in our methodology~\cite{Suspension_rules} as for the following search recommendation and amplification of confusion analysis the results would not include those related to these accounts.
A challenge we had to overcome in crawling \replaced[]{\emph{X}}{Twitter} stems from API request rate limits which the platform imposes to prevent denial of service (DoS). One could alternatively web crawl the platform using multiple registered accounts. Such an approach, has ethical concerns since it imposes a strain on the studied network while it is also less predictable and reproducible since the client side structure of the network might change at any point. In our study, we leverage \replaced[]{\emph{X}}{Twitter}'s \deleted[]{newly released ($2021$) }academic research product~\footnote{Academic Research Access was deprecated on March 2023~\cite{Twitter_Academic_API_Deprecated} but the full-archive search is still available through a paid \emph{pro} or \emph{enterprise} level account.} track authorization~\cite{Twitter_Academic_API} \added[]{($2021$)}. This allows us to freely access the full-archive search endpoint of \replaced[]{\emph{X}}{the Twitter} API (with access to public conversations and tweets posted back to $2006$) with less strict rate limits~\cite{Twitter_Full_Archive}. 

\vspace{3pt}\noindent\textbf{Username Variant Generation.}\label{subsec:username_variant_generation}
\ignore{\vspace{5pt}\noindent\textb{Approach.}} Our work focuses on studying usernames that might be confusingly similar with ones of already popular accounts. Prior works have leveraged seeded network graph traversals for identifying potentially fake or impersonation accounts~\cite{Doppelganger_Bot_Attack, Deep_entity_classification}. Such techniques are only practical when the target accounts are within a small distance from the account they are impersonating, which might not necessarily be the case with all username variants. In contrast, squatting techniques lend themselves well to our problem. Existing tools such as URLCrazy~\cite{URLCrazy} and AppCrazy~\cite{Mobile_App_Squatting} even though not complete, they have been proven successful in generating complex string variations. Nonetheless their string generation models are tailored to their specific problems, that of squatting URLs and mobile app identifiers (package names), respectively. \replaced[]{Hence,}{As a result,} these tools suffer from several limitations because they are restricted to their domains (we refer to them as \textit{prior models}). For example, they produce a large number of strings which are incompatible usernames due to the restrictions that OSNs set for valid usernames. Besides, neither of them supports useful generation patterns, like inserting digits at the beginning or end of a word which would be an easy way for an adversary to automate username variant creations. For clarity we defer a more concrete comparison with prior tools to Section~\ref{subsec:competency_of_tools}.

To address the limitations of existing tools, we develop an end-to-end tool called \textbf{\systemName{}} to drive our experiments. \systemName{} takes as input a set of account usernames (the \textit{Initial Seed}). For each account it employs $10$ string generation models (we refer to them as \textit{primitive models}) to produce valid username variants. For each variant, the generation model runs until the username reaches the maximum number of allowed characters~\cite{Username_policy}. The produced variants (‘\textit{Generated Seed}’) are used in our measurement studies. \systemName{} generation models are categorized as follows a) Insertion models, b) Deletion models, and c) Misspellings. Figure~\ref{fig:Squatting_techniques_graph} illustrates this taxonomy with examples. In Appendix~\ref{appendix:gen_models} we briefly describe each model.


\systemName{} introduces
 several important enhancements to improve coverage of username variants compared to existing tools~\cite{URLCrazy,Mobile_App_Squatting}. Firstly, it employs four new generation models, important to username squatting, namely \textit{number insertion}, \textit{number deletion}, \textit{underscore insertion} and \textit{underscore deletion} (see the \emph{blue-shaded} boxes in Figure~\ref{fig:Squatting_techniques_graph}). Secondly, each model is applied repeatedly (\textit{model self-repetition}) on a username. For example, this allows \systemName{}'s \textit{Double Character Insertion} model to uniquely produce ‘@Jimmmyfalllon’. Thirdly, it supports \textit{model stacking}: after applying a model on a username it takes all the generated variants and applies a second model on them. For example, the combination of \textit{Vowel Insertion} and \textit{Vowel Character Substitution} uniquely generates `@BearackObama', `@BoarackObama' etc. We note that i) during \textit{model stacking} the models are \emph{self-repeated} and ii) the generated usernames of the \textit{primitive models} are not reproduced when \textit{model self-repetition} is applied. We measured the contribution of each of the new features of \systemName{} to account discovery and compare it with what can be achieved with \textit{prior models}~\cite{URLCrazy,Mobile_App_Squatting}. The results are presented in Table~\ref{tab:models_coverage}. In Section~\ref{subsec:model_effectiveness} we perform a more detailed analysis for each of \systemName{}'s models.

\vspace{-1mm}

\begin{table}[!htb]
\centering
\caption{Account discovery comparison between prior string generation models, and \systemName{}'s primitive models, primitive models with self-repetition, and primitive models with model stacking.}
\label{tab:models_coverage}
\resizebox{\columnwidth}{!}{%
\begin{tabular}{ccccc}
\hline
\textbf{Accounts} & \multicolumn{1}{l}{\textbf{Prior Models}} & \textbf{Primitive Models} & \textbf{Self-Repetition} & \textbf{Model Stacking} \\ \hline
\textbf{Active} & $2,055$ & $3,416$ & $24,465$ & $13,512$ \\
\textbf{Suspended} & $436$ & $676$ & $6,656$ & $3,029$ \\ \hline
\end{tabular}}
\end{table}

\begin{figure*}[h!]
\centering
\includegraphics[width=0.96\textwidth,height=\textheight,keepaspectratio]{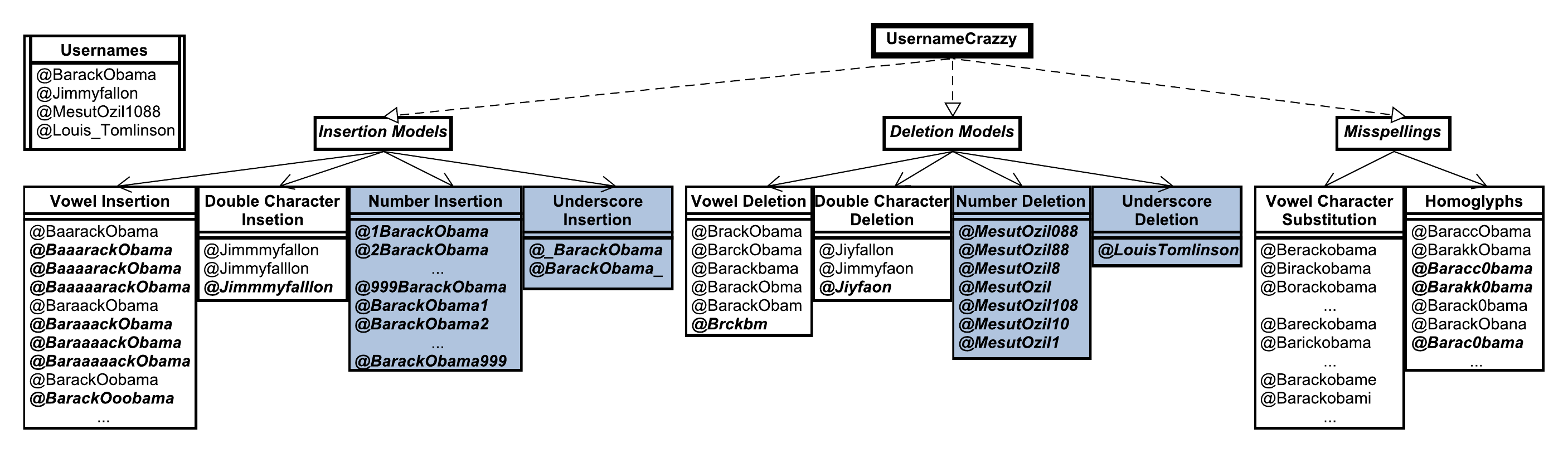}
\captionof{figure}{Taxonomy of \systemName{}'s generation models. The \emph{blue-shaded} boxes indicate models which are not supported by existing tools. In \emph{boldface} are username examples which cannot be generated by existing tools.
\label{fig:Squatting_techniques_graph}}
\vspace{-10pt}
\end{figure*}
%


\subsection{Measurement Methods.}
\subsubsection{\textbf{MQ1.} Measuring Prevalence}
\label{subsec:meth_mq1}
To answer our first MQ, we apply \systemName{} on the \textit{Initial Seed} with the most popular \replaced[]{\emph{X}}{Twitter} accounts~\cite{kaggle_dataset_twitter} ($n=97$) to generate their username variants. We focus on popular accounts since these are more likely to be targeted by username squatting\ignore{\tl{what about this assumption? we do not say why this can be indeed true}}. Three independent raters went through the accounts and labelled them according to \replaced[]{\emph{X}}{Twitter}'s accepted account types~\cite{Twitter_Badge}: a) \emph{‘Government’} ($5$ accounts), b) \emph{‘Companies, brands and organizations’} ($5$ accounts), c) \emph{‘News organizations and journalists’} ($10$ accounts), d) \emph{‘Entertainment’} ($59$ accounts), e) \emph{‘Sports and gaming’} ($15$ accounts) and f) \emph{‘Activists, organizers and other influential individuals’} ($3$ accounts). Conflicts were resolved with discussion.\ignore{The exact categorization can be found in Appendix \ref{apppendix:user_classification}.} Then we search for each of the variants on \replaced[]{\emph{X}}{the social network } to find how many of them currently exist. We call such accounts \textit{active}. To better understand if such username squatting can be exploited for malicious purposes, we use \replaced[]{\emph{X}}{Twitter}'s Academic API v$2$ (\texttt{User Lookup}~\cite{Twitter_User_Lookup}) to check if any of the generated variants were already suspended. 

\subsubsection{\textbf{MQ2.} Measuring Squatted Account Effects}
\label{subsec:meth_mq2}
To understand the impact of squatted accounts we focus on features of OSNs that allow users to interact with and find other users. These are likely to be affected by username squatting issues since they depend on users typing usernames of other users and any typos made will drive traffic to and amplify the impact of squatted accounts. In particular, we focus on the ability of users to \emph{mention} and to \emph{search} for other users. Both are popular features present in most OSNs.

\vspace{3pt}\noindent\textbf{User Mentions.}
\label{subsec:user_mentions}
We conduct experiments to measure whether username squatting affects the \emph{mention} functionality. More interesting to our analysis are \textit{typo-mentions}. We define a \textit{typo-mention} as a mention of a username which was likely made by mistake. Such typo-mentions can drive more activity to the squatted account, amplifying their potentially malicious impact. To measure such typo-mentions we use the full-archive search endpoint (API v$2$~\cite{Twitter_Full_Archive}) to search for the most recent $500$ tweets that \emph{mentioned} each username variant. These tweets can be \textit{actual tweets}, \textit{retweets} or \textit{replies} and for each tweet, we collect the number of likes, retweets, quoted retweets and whether it can contain sensitive information such as links. Note that a \emph{mention} of a username variant on its own does not necessarily indicate a typo-mention.

To further distinguish typo-mentions from purposeful mentions, we use the following observation: a user is more likely to have made a typo-mention, if the mentioner--mentionee are not related. Or conversely, if they are related, then a mention of a username variant was most likely on purpose. We can determine this relationship by looking at the distance between mentioner-mentionee in the social graph. In our experiments we set this distance to ‘$1$' due to practical limitations such as rate limits imposed by \replaced[]{\emph{X}}{the Twitter} API.

To evaluate how accurately the above method can capture typo-mentions we can randomly select a subset of the \textit{actual tweets} and manually label them as \textit{typo--mentions} and \textit{purposeful--mentions} to create a ground truth set. Then we can apply our method which classifies a tweet as a \textit{typo--mention} if the mentioner and mentionee are within one hop in the social graph or as a \textit{not--typo--mention} otherwise to create a prediction set. An issue with random sampling is that we might end-up with an unbalanced dataset---the majority of the tweets could be cases of \textit{purposeful--mentions}. To overcome this and collect a perfectly balanced dataset, we use purposive sampling~\cite{purp_sample1, purp_sample2} instead: we randomly collect $100$ \textit{actual tweets} that are likely \emph{typo-mentions (i.e where the mentioner and mentionee are \emph{not} friends)} and another $100$ \textit{actual tweets} where the mentioner and mentionee are friends. This process resulted in $200$ sampled tweets covering tweets from $80\%$ of the users in our \textit{Initial Seed}. Another issue we faced was that during labelling of the tweets we found that a number of the tweets are difficult to label. Therefore, we introduce a third ground truth label, '\textit{difficult--to--declare}'. After labeling, the ground truth resulted in $51\%$ \textit{typo-mentions}, $38.5\%$ \textit{purposeful--mentions}, and $10.5\%$ \textit{difficult--to--declare}. We define a \textit{true positive} as a tweet that is marked as \textit{typo-mention} from both our method and the ground truth; false positive is a tweet that our method outputs as a \textit{typo-mention} but in the ground truth is labelled otherwise; a true negative is a tweet that our method outputs as a \textit{not--typo--mention} and in the ground truth that same tweet is labelled as either \textit{purposeful--mention} or \textit{difficult--to--declare}; and a false negative is a tweet that our algorithm outputs as a \textit{not--typo--mention} and in the ground truth that same tweet is labelled as \textit{typo--mention}. The results are summarized on Table~\ref{tab:typoMentionDetectionEvaluation}. These correspond to $74$/$26$/$72$/$28$/$74$/$72.5$ for TP/FP/TN/FN/Precision/Recall. Overall we observe that our  method for detecting typo--mentions, even though simple, it can yield good results ($>70$ on both precision and recall), which allows us to apply it to get an overall understanding of typo--mention prevalence on \replaced[]{\emph{X}}{Twitter}.

\vspace{-5pt}

\begin{table}[H]
\caption{Evaluation of \textit{Typo-Mention} measurement method.}
\label{tab:typoMentionDetectionEvaluation}
\resizebox{\columnwidth}{!}{%
\begin{tabular}{ccccc}
\hline
 &  & \multicolumn{3}{c}{\textbf{Actual Values}} \\ \cline{3-5} 
 &  & \textbf{Typo-Mention} & \textbf{Purposeful-Mention} & \textbf{Difficult-to-Declare} \\ \hline
\multicolumn{1}{c|}{\multirow{2}{*}{\textbf{\begin{tabular}[c]{@{}c@{}}Predicted \\ Values\end{tabular}}}} & \textbf{Typo-Mention} & $74$ & $16$ & $10$ \\
\multicolumn{1}{c|}{} & \textbf{Not-Typo-Mention} & $28$ & $61$ & $11$ \\ \hline
\end{tabular}%
}
\vspace{-3pt}
\end{table}

\vspace{-3mm}

\vspace{3pt}\noindent\textbf{User Search Recommendations. }
Most OSNs make recommendations to users trying to either \textit{mention}, \textit{tag} or \textit{search} for another user. We design experiments to gain evidence on whether such recommendations can exacerbate confusion due to username squatting. In particular, we use the \texttt{GET users/search}~\cite{Twitter_Users_Search} method of \replaced[]{\emph{X}}{the Twitter} API, which takes as input a prefix of a username and returns the first $1000$ matching results. We apply this request for each of the possible prefixes of a username (if a username has $n$ characters, we issue $n$ requests). We repeat this experiment for each\ignore{ account} user of the \textit{Initial Seed} to examine whether the generated\ignore{username} variants are ranked higher in the list of the recommended users than the respective original accounts and how often this phenomenon happens.


\subsubsection{\textbf{MQ3.} Username Variants Characteristics}
\label{subsec:meth_mq3}
For MQ$3$, we aim to analyze the characteristics of squatted usernames to help us understand which ones are more prevalent or contribute to on the most. String generation models used in squatting methodologies can produce string variants that are significantly different from the seeded input but also among themselves. These can change the phonological properties of usernames, but we hypothesize that smaller changes can be more ng as they can result in similarly sounding and similarly looking usernames. To measure this, we\ignore{ calculate} compute the edit distance (ED) between a squatted user and its seeded one using a well--established string similarity metric~\cite{Levenshtein_distance}.

\vspace{-1em}
\vspace{+2.5mm}
\subsubsection{\textbf{MQ4.} Squatted Account Behavior}
\label{subsec:meth_mq4}
To understand whether squatted accounts exhibit suspicious behavior we aim to identify i) how many accounts out of the generated ones have already been suspended, ii) how many accounts are bot accounts, iii) the behavior of squatted accounts that share similar profile picture and name.

\vspace{3pt}\noindent\textbf{Suspended Accounts Analysis.} We use the Academic API v$2$ and we apply the \texttt{User Lookup}~\cite{Twitter_User_Lookup} method on every generated username. This call returns useful static information of an individual or a group of accounts, such as the \textit{name}, \textit{username}, and the \textit{creation date}. If an account is suspended, we conclude that these accounts were involved in some form of malicious (e.g., impersonation) or unwanted (e.g., on) behavior~\cite{Rules_and_policies}. As these accounts no longer exist on the network, \replaced[]{\emph{X}}{the Twitter} API does not return any output for a suspended username. To find the content of those profiles we instead used the \textit{Wayback Machine}~\cite{Wayback_Machine}; an online archive that maintains snapshots of \replaced[]{\emph{X}}{the Twitter}\footnote{The snapshots for which we get a ‘\textit{Got an HTTP 302 response at crawl time}’ error are excluded from the analysis.}. Two independent raters then analyze the recovered profiles. We based the content analysis on \replaced[]{\emph{X}}{Twitter}'s prohibited activity~\cite{Twitter_Botometer_Critique} and classified the profiles as i) impersonation attempts; ii) bots engaged in automated or spamming activity; iii) accounts that explicitly mention they are fan\replaced[]{/parody/}{ accounts, a parody or }fake version of the celebrity account; this kind of ‘impersonation' is allowed by \replaced[]{\emph{X}}{Twitter}'s policy~\cite{Twitter_policies}; iv) benign users, i.e. users who are not related to the celebrity account except for sharing a username variant and v) other accounts which do not fall in the above categories. 

\vspace{3pt}\noindent\textbf{Active Accounts Analysis.}
We first run Botometer~\cite{Botometer} for \emph{every} generated variant to understand how many of the accounts are bots. Botometer supports two types of scores, ‘\textit{English}’ and ‘\textit{Universal}’ (language independent). The former takes into account $6$ categories of features, while the latter omits $2$ of them (content and sentiment features) which are English-specific. Note that the ‘\textit{Universal}’ score is not as accurate as the ‘\textit{English}’~\cite{Botometer_Universal_model, Botometer_Universal_model_2}. Consequently, during our experiments we try to work with English-related users as much as possible. We note that we examine only a subset of our users because Botometer does not return a score for locked accounts. \ignore{or for profiles with no activity.}

To gain a deeper understanding into the activity of the active profiles, we perform a manual analysis of the activity of $1,400$ randomly selected squatted username accounts that have a profile picture with a face or an avatar. Based on previous works~\cite{profile_photos_1, profile_photos_2}, profile pictures with faces are more persuasive in terms of influencing user behavior, engagement or credibility compared to pictures with other objects. Two independent raters then analyse the behavior of the accounts. If an account had no suspicious behavior it was labeled as \emph{benign}. If an account can contribute to on we term it \emph{suspicious}; 
these were subdivided into the following: a) \textit{impersonation}: an account that has no specific goal other than impersonate, e.g., this impersonation could be done for entertainment purposes or the creator of the account could be a fan, but does not explicitly mention so on the account~\cite{Twitter_policies}; b) \textit{financial} accounts that use the generated on for a financial incentive, e.g. they try to sell products or gain followers; c) \textit{political} aiming to use the on to influence their followers politically, e.g., such accounts may (re)post news or opinion articles; d) \textit{news} accounts that spread (fake) news about the celebrity they impersonate, these accounts differ from the impersonation accounts as they have the specific goal of spreading news about the seed account holder; e) \textit{harass}: accounts that directly harass (e.g., insult or aggressively mention or retweet) the celebrity they impersonate or other celebrities; and f) \textit{other}: other accounts that did not fall into the other categories, e.g., accounts with little or unrelated activity.

\vspace{3pt}\noindent\textbf{Amplification of Confusion.}\label{subsec:amplification_of_confusion}
Important to our analysis is to understand whether squatted accounts with malicious purposes exploit other similarities with the original account.
To measure this we search for active squatted accounts with similar \textit{profile names} and \textit{\ignore{profile} images} to the original and analyze how many of those accounts are bots. We apply VGGFace$2$~\cite{VGGFace2}, a popular image recognition model (see Section~\ref{subsec:features} and Algorithm~\ref{alg:Image_Profile_Similarity} in Appendix~\ref{appendix:image_similarity_algorithm} for details of the algorithm) on all squatted usernames~\footnote{Profile images that do not depict a face are excluded.} and then we compare, using the Levenshtein distance~\cite{Levenshtein_distance}, their profiles names with their potential target. We consider profile names to be similar if either i) two names are equivalent, ii) an extra character is added as a prefix or suffix, iii) a substring of the original name appears as a substring in the squatted name (e.g., for ‘Cristiano Ronaldo’ we \deleted[]{will }check \deleted[]{separately }whether ‘Cristiano’ or ‘Ronaldo’ appear in the squatted account's profile name). Details of the algorithm can be found in \deleted[]{the }Appendix~\ref{appendix:profile_name_similarity_algorithm}, Algorithm~\ref{alg:Profile_Name_Similarity}. Out of those\ignore{ accounts} users which share similar image and/or name, we compute how many are bots using Botometer~\cite{Botometer}.

%% file: sections/4_results.tex
\section{Characterization}
\label{sec:results}


\subsection{Prevalence of Username Squatting.}
\label{sec:prevalence}

We conducted our \textbf{MQ$\textbf{1}$} measurements on \replaced[]{\emph{X}}{Twitter}, in October $2021$ and found a total of $41,393$ active username variants from the \emph{Initial Seed} ($n=97$) of popular accounts, or approximately $427$ variants per username. Table~\ref{tab:Table 1} lists the $5$ top original usernames with the most active username variants. We also found that $10,361$ accounts with username variants were suspended, or $107$ on average per original account. This \replaced[]{is a good signal}{indicate} that in the past these accounts were malicious~\cite{Rules_and_policies} which we further analyze in Section~\ref{subsec:suspended_analysis}. Next, we examined how these suspended accounts are spread across different categories. We found that $7,940$ accounts ($66.3\%$) belong to the ‘Entertainment' category, $1,175$ accounts ($9.8\%$) in ‘Sport and Gaming', $1,104$ accounts ($9.2\%$) in ‘Companies, Brands and Organizations', $709$ accounts ($5.9\%$) in ‘Government', $582$ accounts ($4.8\%$) in ‘Activists, Organizers and Other Influential Individuals' and $460$ accounts ($3.8\%$) in ‘News Organizations and Journalists' category. In May $2022$, we ran again the measurements and found $43,349$ active and $11,384$ suspended accounts while in November $2022$ we found $43,994$ active and $11,827$ suspended accounts (see Table~\ref{tab:Table 1}). Alarmingly, the increase on the number of the suspended\ignore{ accounts} users in every new snapshot\ignore{ demonstrates} indicates that this is an ongoing and growing problem. For the rest of the paper, we use the first snapshot. \ignore{Therefore}Thus, we conclude that celebrities and popular\ignore{ accounts} users in general are indeed a common target of username squatting attack, something that contrasts the insights of previous work which found\ignore{only} $166$ potential\ignore{ impersonation pairs} impersonators and only $3$ of them were celebrity impersonation attempts~\cite{Doppelganger_Bot_Attack} (see Sections~\ref{subsec:squad_popular} and~\ref{sec:background}).

\begin{table}[!ht]
\caption{Top 5 original accounts with the most active username variants on \replaced[]{\emph{X}}{Twitter}.}
\label{tab:Table 1}
\setlength{\tabcolsep}{1pt}
\resizebox{\columnwidth}{!}{%
\begin{tabular}{lcccccc}
\hline
\multicolumn{7}{c}{\textbf{Potential Squatted Usernames}} \\ \hline
\multicolumn{1}{c}{\textbf{\begin{tabular}[c]{@{}c@{}}Original\\ Accounts\end{tabular}}} &
  \multicolumn{1}{c}{\textbf{\begin{tabular}[c]{@{}c@{}}Generated \\ Accounts\end{tabular}}} &
  \multicolumn{1}{c}{\textbf{\begin{tabular}[c]{@{}c@{}}Gen. Accts. \\ with ED $\textbf{1}$-$\textbf{3}$\end{tabular}}} &
  \multicolumn{1}{c}{\textbf{\begin{tabular}[c]{@{}c@{}}Active\\ Accounts\end{tabular}}} &
  \multicolumn{1}{c}{\textbf{\begin{tabular}[c]{@{}c@{}}Act. Accts. \\ with ED $\textbf{1}$-$\textbf{3}$\end{tabular}}} &
  \multicolumn{1}{c}{\textbf{\begin{tabular}[c]{@{}c@{}}Suspended\\ Accounts\end{tabular}}} &
  \textbf{\begin{tabular}[c]{@{}c@{}}Sus. Accts. \\ with ED $\textbf{1}$-$\textbf{3}$\end{tabular}} \\ \hline
\multicolumn{1}{l}{@kaka} &
  \multicolumn{1}{c}{$5,899$} &
  \multicolumn{1}{c}{$1,380$} &
  \multicolumn{1}{c}{$2,145$} &
  \multicolumn{1}{c}{$441$} &
  \multicolumn{1}{c}{$224$} &
  $96$ \\ 
\multicolumn{1}{l}{@cristiano} &
  \multicolumn{1}{c}{$17,687$} &
  \multicolumn{1}{c}{$2,301$} &
  \multicolumn{1}{c}{$1,714$} &
  \multicolumn{1}{c}{$639$} &
  \multicolumn{1}{c}{$148$} &
  $63$ \\ 
\multicolumn{1}{l}{@pink} &
  \multicolumn{1}{c}{$11,955$} &
  \multicolumn{1}{c}{$1,348$} &
  \multicolumn{1}{c}{$1,555$} &
  \multicolumn{1}{c}{$381$} &
  \multicolumn{1}{c}{$162$} &
  $75$ \\ 
\multicolumn{1}{l}{@nasa} &
  \multicolumn{1}{c}{$5,842$} &
  \multicolumn{1}{c}{$1,356$} &
  \multicolumn{1}{c}{$1,534$} &
  \multicolumn{1}{c}{$333$} &
  \multicolumn{1}{c}{$231$} &
  $42$ \\ 
\multicolumn{1}{l}{@selenagomez} &
  \multicolumn{1}{c}{$9,925$} &
  \multicolumn{1}{c}{$2,270$} &
  \multicolumn{1}{c}{$1,346$} &
  \multicolumn{1}{c}{$306$} &
  \multicolumn{1}{c}{$177$} &
  $97$ \\ \hline
\multicolumn{1}{c}{\textbf{\begin{tabular}[c]{@{}c@{}}Total  (Oct. $\textbf{2021}$)\end{tabular}}} &
  \multicolumn{1}{c}{\multirow{3}{*}{$851,682$}} &
  \multicolumn{1}{c}{\multirow{3}{*}{$378,088$}} &
  \multicolumn{1}{c}{\begin{tabular}[c]{@{}c@{}}$41,546$\end{tabular}} &
  \multicolumn{1}{c}{\begin{tabular}[c]{@{}c@{}}$37,965$\end{tabular}} &
  \multicolumn{1}{c}{\begin{tabular}[c]{@{}c@{}}$10,208$\end{tabular}} &
  \begin{tabular}[c]{@{}c@{}}$9,762$\end{tabular} \\ 
\multicolumn{1}{c}{\textbf{\begin{tabular}[c]{@{}c@{}}Total  (May $\textbf{2022}$)\end{tabular}}} &
  \multicolumn{1}{c}{} &
  \multicolumn{1}{c}{} &
  \multicolumn{1}{c}{\begin{tabular}[c]{@{}c@{}}$43,349$\end{tabular}} &
  \multicolumn{1}{c}{\begin{tabular}[c]{@{}c@{}}$39,403$\end{tabular}} &
  \multicolumn{1}{c}{\begin{tabular}[c]{@{}c@{}}$11,384$\end{tabular}} &
  \begin{tabular}[c]{@{}c@{}}$10,781$\end{tabular} \\ 
\multicolumn{1}{c}{\textbf{\begin{tabular}[c]{@{}c@{}}Total  (Nov. $\textbf{2022}$)\end{tabular}}} &
  \multicolumn{1}{c}{} &
  \multicolumn{1}{c}{} &
  \multicolumn{1}{c}{\begin{tabular}[c]{@{}c@{}}$43,994$\end{tabular}} &
  \multicolumn{1}{c}{\begin{tabular}[c]{@{}c@{}}$40,200$\end{tabular}} &
  \multicolumn{1}{c}{\begin{tabular}[c]{@{}c@{}}$11,827$\end{tabular}} &
  \begin{tabular}[c]{@{}c@{}}$11,133$\end{tabular} \\ \hline
\end{tabular}}
\vspace{-10pt}
\end{table}

\subsection{Online Confusion.} \label{sec:online_confusion}
\noindent\textbf{User Mentions.} For \textbf{MQ$\textbf{2}$}, by applying the methodology described in Section~\ref{subsec:meth_mq2} we collected $1.2$\ignore{$1,232,746$} million tweets to measure the contribution of squatted accounts to online confusion. Tweets can either be \textit{actual tweets}, \textit{retweets}, or \textit{replies} that mention at least one of the generated username variants. We found that out of the $864,369$ actual tweets, $657,337$/$864,369$ ($76\%$) are likely \emph{typo-mentions} and only $200,562$/$864,369$ ($23.2\%$) happen when the mentioner and mentionee are within one hop on the social graph (\textit{purposeful--mentions}). We discard the rest of the tweets ($6,470$ or $0.8\%$) because the API did not return information about the friendship of the users.


We further analyze the $657,337$ \emph{typo-mentions} and\ignore{(Figure~\ref{fig:Twitter_actual_tweets_per_category})} found that $437,550$ tweets ($66.6\%$) belong to the ‘Entertainment' category, $96,720$ tweets ($14.7\%$) in ‘Sport and Gaming', $57,126$ tweets ($8.7\%$) in ‘Companies, Brands and Organizations', $33.220$ tweets ($5.1\%$) in ‘News Organizations and Journalists' category, $25,206$ tweets ($3.8\%$) in ‘Government' and $7,515$ tweets ($1.1\%$) in ‘Activists, Organizers and Other Influential Individuals'. We observe that typo-mentions are prevalent in all categories but most are found in Entertainment accounts as expected since our dataset contains more such accounts. Typo-mentions are frequent which is worrisome since they can drive more traffic to the suspicious account.

\vspace{+1mm}
\ignore{
\subsection{Online confusion.} \label{sec:online_confusion}
\noindent\textbf{User Mentions.} For MQ$2$, we collected $1.2$\ignore{$1,232,746$} million tweets to measure the contribution of squatted accounts to online confusion. Tweets can either be \textit{actual tweets}, \textit{retweets}, or \textit{replies} that mention at least one of the generated username variants. We found that out of the $864,369$ actual tweets, $657,337$/$864,369$ ($76\%$) are likely \emph{typo-mentions}\tl{; we call this group \emph{typos}} and only $200,562$/$864,369$ ($23.2\%$) happen when the mentioner and mentionee are within one hop on the social graph\tl{; we call this group \emph{no\_typos}}. \tl{We discard the rest $6,470$ ($0.8\%$) tweets because the API did not return information about the friendship of the users. To validate our hypothesis we use a purposive sample~\cite{purp_sample1, purp_sample2} of $200$ tweets; we randomly select $100$ tweets from \emph{typos} group and another $100$ from the \emph{no\_typos} ? TODO (PASTE THE ANSWER HERE) randomly select $100$ tweets from the tweets that there is no friendship and another 100 tweets from the other team (covering $78$ out of $97$ $(55.3\%)$ users in total of our \textit{Initial Seed}). For the first $100$ tweets: i) $62$ $(62\%)$ tweets marked as \emph{typo-mentions}, ii) $24$ $(24\%)$ as \emph{no typo-mentions} and iii) $14$ $(14\%)$ as ‘difficult to declare'. For the other $100$ tweets: $34$ $(34\%)$ marked as \emph{typo-mentions}, ii) $57/100$ $(57\%)$ as \emph{no typo-mentions} and iii) $9/100$ $(9\%)$ as ‘difficult to declare'. We further analyze \emph{mentions} with \emph{no relationship} of the users per account category and\ignore{(Figure~\ref{fig:Twitter_actual_tweets_per_category})} notice that $437,550$ tweets ($66.6\%$) belong to the ‘Entertainment' category, $96,720$ tweets ($14.7\%$) in ‘Sport and Gaming', $57,126$ tweets ($8.7\%$) in ‘Companies, Brands and Organizations', $33.220$ tweets ($5.1\%$) in ‘News Organizations and Journalists' category, $25,206$ tweets ($3.8\%$) in ‘Government' and $7,515$ tweets ($1.1\%$) in ‘Activists, Organizers and Other Influential Individuals'. We observe that typo-mentions are prevalent in all categories but most are found in Entertainment accounts as expected since our dataset contains more such accounts. Typo-mentions are frequent which is worrisome since they can drive more traffic to the suspicious account.}
}

\vspace{3pt}\noindent\textbf{User Search Recommendations.}\label{subcsec:user_search_recommendations} In order to identify the factors that can lead to the increased popularity of potential squatted accounts, we take a closer look at the suggestion algorithm of \replaced[]{\emph{X}}{Twitter}. We expect that when a user searches for a celebrity using the search bar the targeted profile should be returned within the first $10$ options due to its popularity. Using the methodology from Section \ref{subsec:meth_mq2}, out of the $929$ requests (the total number of the characters of our \textit{Initial Seed}) we find that only $166$ and $178$ times all the original users together appear in the first $10$ and $100$ recommendations, respectively. Alarmingly, only when the full original usernames were inserted in the search bar, the official profiles were returned as the first $15$ options (see Table~\ref{tab:original_users_recommendation}). 
We observed that there exist instances where the squatted username appears \emph{before} the original account\deleted[]{(example on Figure~\ref{fig:suggested_user} in Appendix~\ref{appendix:more_examples})}. Clearly, this is problematic as it can exacerbate the problem of online confusion\ignore{ since}.
More precisely, $1,205$ of active squatted accounts appear in the first $1,000$ recommendations, $705$ of them appear in the top $500$ recommendations, $170$ in the top $100$ and $26$ appear in the top $10$ recommendations while searching for our seed accounts (see Figure~\ref{fig:Twitter_recommended_users_with_edit_distance}). 
This indicates that the popularity of squatted usernames can be amplified through the search algorithm of the platform itself. Moreover, we manually check the $170$ squatted \ignore{accounts}users appearing in the top $100$ recommendations and marked $10\%$\ignore{17 users} as impersonators and $38.8\%$\ignore{49 + 17 = 66 users} as malicious. This shows alarmingly that search recommendations can not only amplify confusion but can also be leveraged by malicious actors.

\vspace{+2mm}

\vspace{-2mm}

\begin{table}[!htb]
\centering
\caption{Original accounts within the first 10 and 100 search recommendation results from a total of 929 requests.}
\label{tab:original_users_recommendation}
\resizebox{\columnwidth}{!}{%
\begin{tabular}{ccc}
\hline
\multicolumn{3}{c}{\textbf{Original Accounts - \replaced[]{\emph{X}}{Twitter}'s Recommendation System}}                          \\ \hline
\textbf{Position} & \textbf{\begin{tabular}[c]{@{}c@{}}\# of Times\\ Original Accounts\\ Recommended\end{tabular}} & \textbf{Observations} \\ \hline
\multicolumn{1}{c}{\textbf{Top $\textbf{10}$}} &
  \multicolumn{1}{c}{$166$} &
  \begin{tabular}[c]{@{}c@{}}$108$ times the original accounts returned as first\\ recommendation but only when their full\\ name was inserted in the search bar (ED = $0$)\end{tabular} \\ \hline
\multicolumn{1}{c}{\textbf{Top $\textbf{100}$}} &
  \multicolumn{1}{c}{$178$} &
  \begin{tabular}[c]{@{}c@{}} $132$ times the original accounts returned in position \\ $11$ to $15$ but only when their full name\\ was inserted in the search bar (ED = $0$)\end{tabular} \\ \hline
\end{tabular}}
\vspace{-10pt}
\end{table}

\subsection{Squatted Username Characteristics.}
\label{subsec:characteristics}
For \textbf{MQ$\textbf{3}$}, we aim to analyze the characteristics of squatted usernames to\ignore{ help us} \replaced[]{identify}{understand} which ones are more prevalent or contribute to confusion the most. To do that we use the edit distance measure (see Section~\ref{sec:measurement_study}). 
The highest number of suspended and active accounts are squatted usernames where the \replaced[]{\emph{ED}}{edit distance} is between \replaced[]{$1$-$3$}{one and three} characters (see Figure~\ref{fig:Twitter_suspended_existing_edit_distance} in Appendix~\ref{appendix:username_characteristics}). Looking into the high numbers for \emph{$ED=3$}, we found that in numerous cases, username variants consist of an addition of \emph{three} digits before or after the seed username.

\begin{figure}[!ht]
    \centering
    \includegraphics[width=0.91\columnwidth,height=\textheight,keepaspectratio]{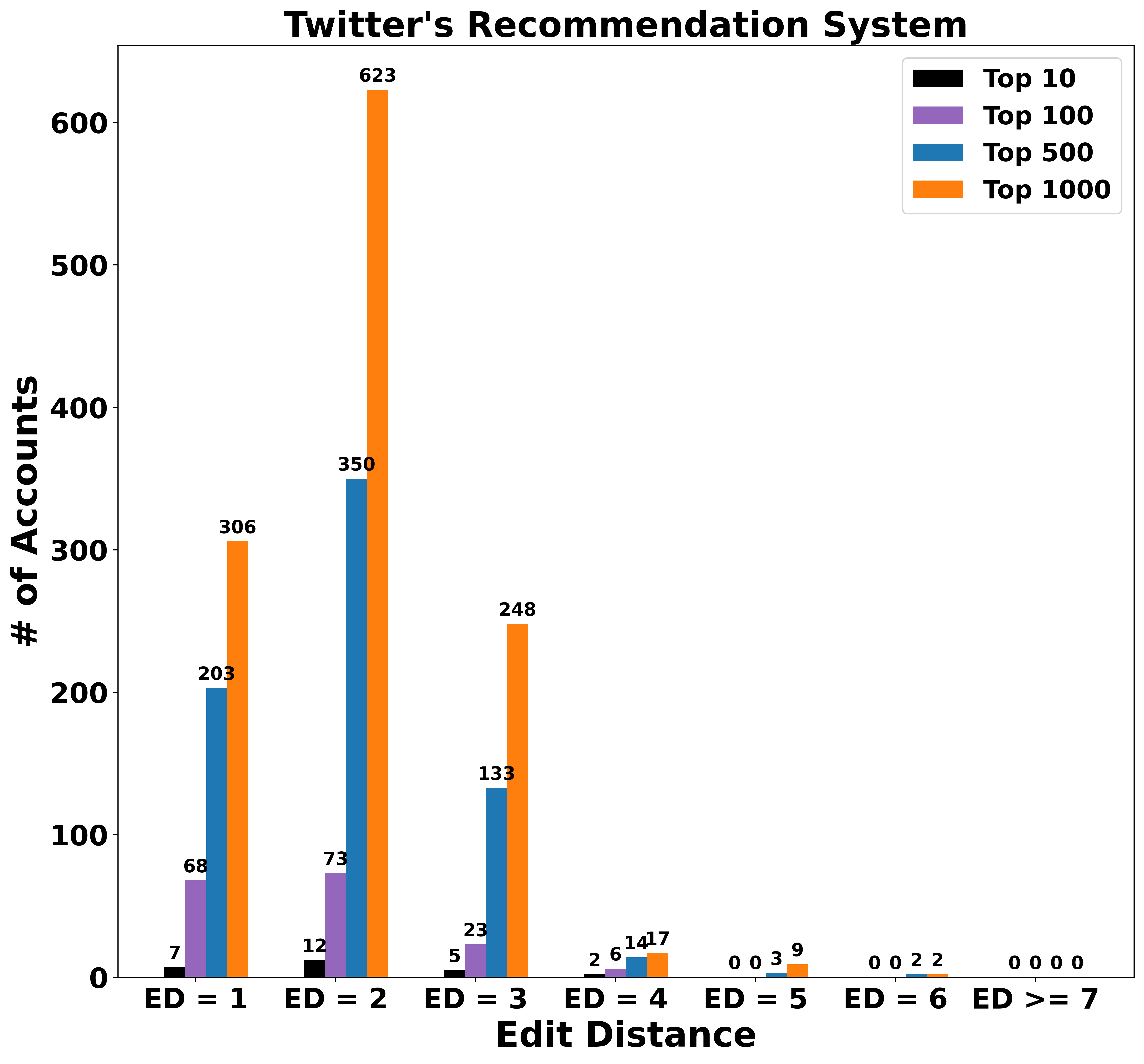}
    \caption{Squatted accounts in the top search recommendations grouped by edit distance.}
    \label{fig:Twitter_recommended_users_with_edit_distance}
\end{figure}

\begin{figure}[!htb]
    \centering
    \includegraphics[width=0.91\columnwidth,height=\textheight,keepaspectratio]{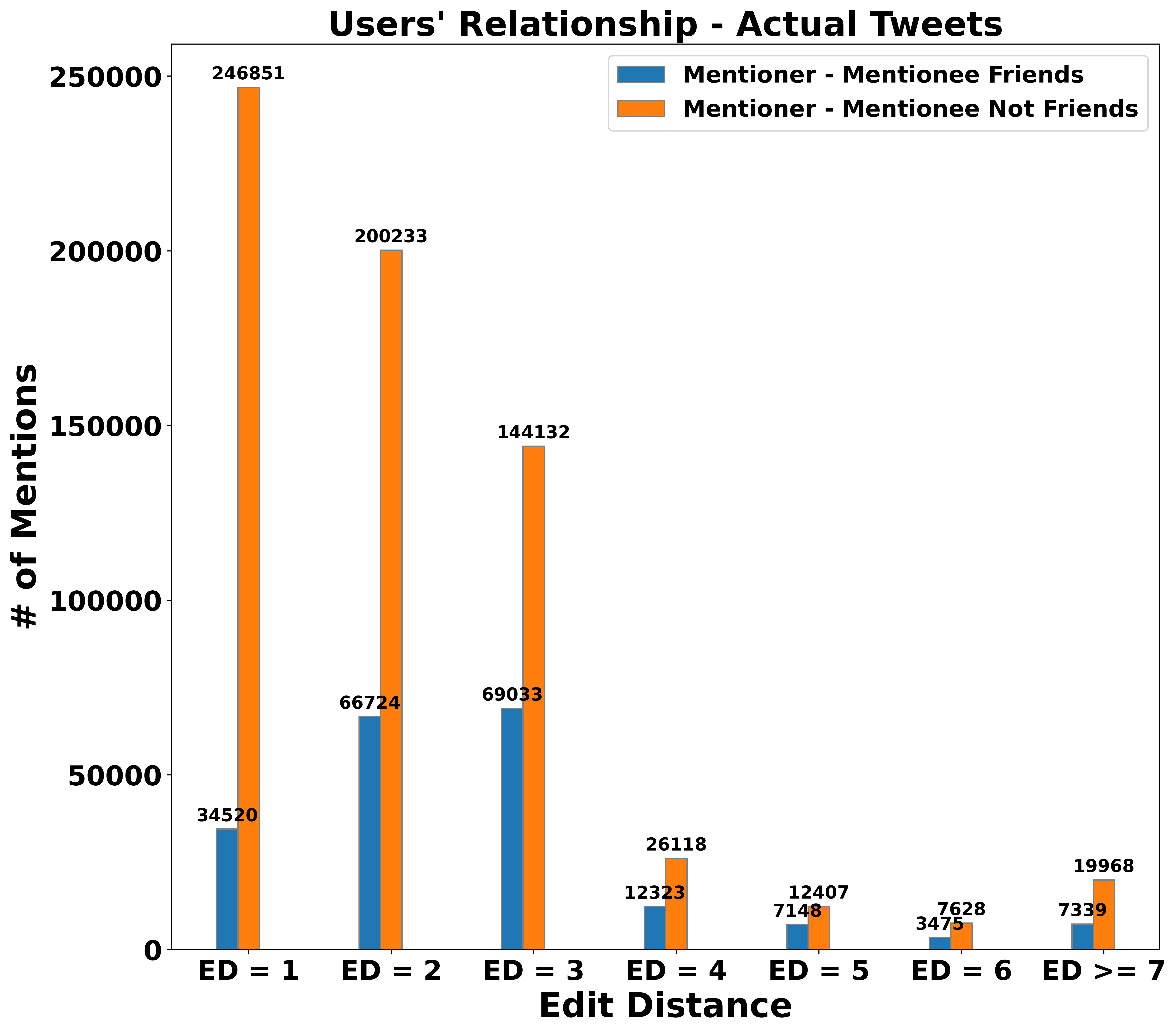}
    \caption{Number of actual tweets that mention at least one username variant grouped by the variants' edit distance with their original accounts.}
    \label{fig:Twitter_edit_distance_actual_tweets}
\end{figure}

Making $1$-$3$ mistakes when typing a word is reasonable~\cite{Typo-mistakes_paper_A,Typo-mistakes_paper_B}. Figure \ref{fig:Twitter_edit_distance_actual_tweets} \deleted[]{further }depicts that i) indeed users often make \replaced[]{$1$-$3$}{$1$ to $3$} mistakes in their mentions and ii) most of the times there is no relationship between the mentioner and the mentionee. The number of mistakes in actual tweets is more informative than in replies or retweets, as the latter can simply propagate the same typo mistakes. Figure~\ref{fig:Twitter_recommended_users_with_edit_distance} illustrates the effects of the edit distance on \replaced[]{\emph{X}}{Twitter}'s search recommendations. Out of the squatted accounts suggested by the search algorithm, again we note that those with \replaced[]{\emph{ED = 1}}{edit distance $1$-$3$} are most common. Interestingly, $24$ squatted\ignore{ usernames} users appear within the top $10$, $164$ within the top $100$ and $686$ within the top $500$ of the search recommendations. From these results we can conclude that an adversary can identify specific patterns of typo mistakes and claim a username that is close to the original. Our results indicate that typo-mentions and the search algorithm can be exploited to increase traffic to squatted accounts.

\subsection{Squatted Account Behavior.}
\label{subsec:activity_of_suspended_and_active_accounts}
For \textbf{MQ$\textbf{4}$}, we aim to analyze the activity of both suspended and active accounts. For this we manually analysed the behavior of $135$ suspended accounts using the Wayback Machine, we classified the behavior of $540$ suspicious and active accounts and we applied Botometer to all active accounts.

\vspace{3pt}\noindent\textbf{Suspended Accounts Analysis.}
\label{subsec:suspended_analysis}
By applying the methodology described on Section~\ref{subsec:meth_mq4} we found $10,208$ suspended accounts. \textit{Wayback Machine}~\cite{Wayback_Machine} returned valid snapshots for $135$ of those accounts\ignore{\footnote{\tl{For the rest \ignore{$10,226$}accounts, Wayback Machine returned no or corrupted snapshots.}}} \ignore{\st{for these accounts}}. Then, two independent raters went through all the available snapshots and they found (see Table~\ref{tab:suspended_raters}): $52$ impersonation attempts of the original account, $12$ accounts posting e.g., spam links or product promotions, $6$ human accounts that explicitly mention they are fan or parody accounts, $15$ profiles of humans with no connection to the original account, $45$ other accounts. Other include private profiles or accounts with no activity. Also, for $5$ accounts there was a disagreement between the raters so we do not report their results. This analysis provides a first validation of our idea that a suspension was the consequence of malicious content creation. 

\begin{table}[!htb]
\centering
\small
\caption{Manual classification of suspended accounts.}
\label{tab:suspended_raters}
\resizebox{\columnwidth}{!}{%
\begin{tabular}{ccccc}
\hline
\multicolumn{5}{c}{\textbf{Suspended Accounts ($\textbf{135}$ users)}} \\ \hline
\multicolumn{1}{c}{\textbf{\begin{tabular}[c]{@{}c@{}}Impersonations \end{tabular}}} &
  \multicolumn{1}{c}{\textbf{Bots}} &
  \multicolumn{1}{c}{\textbf{Fans}} &
  \multicolumn{1}{c}{\textbf{Humans}} &
  \textbf{Other} \\ 
\multicolumn{1}{c}{$52$ ($38.5\%$)} &
  \multicolumn{1}{c}{$12$ ($8.8\%$)} &
  \multicolumn{1}{c}{$6$ ($4.4\%$)} &
  \multicolumn{1}{c}{$15$ ($11.1\%$)} &
  $45$ ($33.3\%$) \\ \hline
\end{tabular}}
\end{table}

\vspace{3pt}\noindent\textbf{Active Accounts Analysis.}
\label{subsec:bot_analysis}
Previous \ignore{research }works~\cite{Why_Botometer_1, Why_Botometer_2} used Botometer as ground truth for measuring the bot activity of accounts. To validate our hypothesis that many of the squatted usernames are indeed malicious and/or bots, we applied Botometer to all the generated usernames that are still \emph{active} on \replaced[]{\emph{X}}{the OSN}. Botometer uses \textit{CAP} score, a probability which indicates that a profile with this score or greater is controlled by a software (i.e., is a bot). We set \textit{CAP} score to $0.95$, which is a conservative value expected to yield only $5\%$ false positives~\cite{Botometer}. Our analysis\deleted[]{( see Appendix~\ref{appendix:botometer_cap_selection} and Figure~\ref{fig:bot_cap} in Appendix~\ref{appendix:botometer_results})} shows that on our dataset this value can be around $11\%$. Out of \textit{$41,546$} variants Botometer returned a score for \textit{$25,089$} since the other accounts are private profiles. \ignore{$9,702$ or $38.7\%$ of the profiles}$9,702$ ($38.7\%$) profiles were classified as automated\ignore{ accounts} users. The automated accounts are further categorized as different kinds of bots and the largest category was \textit{fake followers}\deleted[]{( see Table~\ref{tab:Botometer_types} in Appendix~\ref{appendix:bot_types} and Figure~\ref{fig:bot_types} in Appendix~\ref{appendix:botometer_analysis}}\ignore{for more details)}. We observe that\ignore{ indeed} a large number of the squatted\ignore{ accounts} users are malicious bots, indicating also the effectiveness of \systemName{}.

We next performed a manual analysis (see Section~\ref{subsec:meth_mq1}) to better understand the \emph{type} of activity the squatted accounts are involved in. Out of the $41,546$ active profiles we randomly selected $1,400$ users who have a face or an avatar in their profile picture. Two raters agreed beforehand on the rules they will follow, discussing about the characteristics of impersonators, spam/bot and benign users. Accounts that had no similarities with the seed account or explicitly say they were fan or parody \ignore{accounts}profiles were labelled as \emph{benign}~\cite{Twitter_policies}, while accounts that were sharing same features with their original accounts were labelled as \emph{suspicious}. Also, \ignore{profiles}users sharing a number of features\footnote{Profile name or picture, username.} as their seed account but with limited or no activity were labelled as \emph{suspicious}. The process allowed us to identify $838$ accounts as \textit{benign}, $540$ as \textit{suspicious} (defined as an account that can contribute to confusion) username squatting attempts and $22$ accounts were left as ‘difficult to declare'. We examined the inter-rater agreement using Cohen'a Kappa~\cite{Cohen’a_kappa} and found $\kappa=0.92$ and $\kappa=0.89$ for the benign and suspicious users, respectively, which indicates \textit{near perfect} agreement.
Then, the two manual raters further categorized the suspicious accounts' behavior as \textit{impersonation}, \textit{financial}, \textit{political}, \textit{news} and \textit{harass}. The full results for the suspicious accounts are presented in Table~\ref{tab:malicious_users_raters}. We conclude that a large amount of active accounts are involved in some form of malicious or confusing behavior.

\vspace{-1.2mm}
\begin{table}[!htb]
\centering
\small
\caption{Manual classification of the active malicious accounts used in dataset.}
\label{tab:malicious_users_raters}
\resizebox{\columnwidth}{!}{%
\begin{tabular}{cccccc}
\hline
\multicolumn{6}{c}{\textbf{Active Malicious Accounts ($\textbf{540}$ users)}} \\ \hline
\multicolumn{1}{c}{\textbf{\begin{tabular}[c]{@{}c@{}}Impersonations \end{tabular}}} &
  \multicolumn{1}{c}{\textbf{Financial}} &
  \multicolumn{1}{c}{\textbf{Political}} &
  \multicolumn{1}{c}{\textbf{News}} &
  \multicolumn{1}{c}{\textbf{Harass}} &
  \textbf{Other\tablefootnote{Accounts with little or unrelated activity.}} \\ 
\multicolumn{1}{c}{$115$ ($21.3\%$)} &
  \multicolumn{1}{c}{$56$ ($10.3\%$)} &
  \multicolumn{1}{c}{$7$ ($1.3\%$)} &
  \multicolumn{1}{c}{$15$ ($2.6\%$)} &
  \multicolumn{1}{c}{$4$ ($0.7\%$)} &
  \multicolumn{1}{l}{$343$ ($63.5\%$)} \\ \hline
\end{tabular}}
\vspace{-5pt}
\end{table}

\vspace{-1mm}

\begin{figure}[!htb]
\centering
\includegraphics[width=0.95\columnwidth,height=\textheight,keepaspectratio]{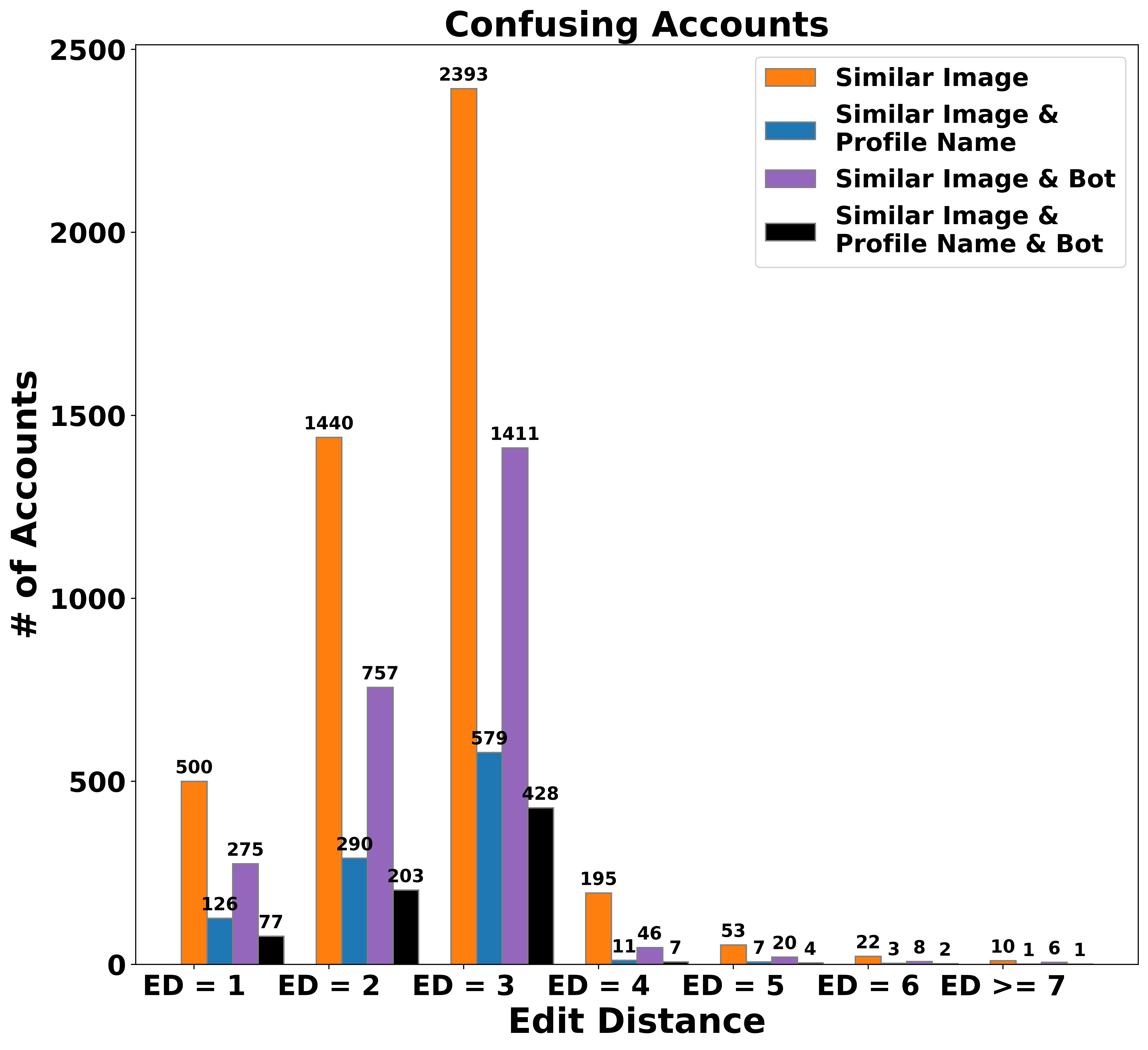}
\caption{Classification of the confusing accounts.}
\label{fig:Confusing_accounts_classification}
\vspace{-10pt}
\end{figure}

\vspace{3pt}\noindent\textbf{Analysing Confusion Amplifiers.}
\label{subsec:Analysing_Confusion_Amplifiers}
Using the method presented in Section \ref{subsec:meth_mq4} we analyze how many bot accounts share similar features (here profile name/image) with the original \ignore{account}one. Figure~\ref{fig:Confusing_accounts_classification} depicts that indeed a \ignore{substantial}large number of users have a similar name/image with their target; this suggests that these profiles likely impersonate the original owner to drive traffic to their account and increase the impact of their malicious activity. \deleted[]{On GitHub~\cite{SQUAD_Github} we show examples of fake tweets and suspected bot accounts posting suspicious link with their profile.} 
\ignore{In Appendix~\ref{appendix:more_examples} we show an example of a fake tweet (Figure~\ref{fig:fake_trump}) and a suspected bot account posting a suspicious link (Figure~\ref{fig:malicious_links_impersonation}) with its profile (Figure~\ref{fig:malicious_links_profile}).}

\vspace{+2mm}

%% file: sections/5_design.tex
\section{\frameworkName{} Design Overview}
\label{sec:design}
Our measurement study revealed that username squatting is an important issue on \replaced[]{\emph{X}}{Twitter} as a significant subset of these accounts are bot\replaced[]{s}{ accounts} or impersonations involved in malicious behavior. On the other hand, a large number of squatted \replaced[]{users}{accounts} are benign (e.g., \replaced[]{fan/parody users}{fan and parody accounts}) and we need an effective way to make that distinction. This raised the following question: \textit{Can we design an effective and efficient framework for detecting suspicious username squatting attempts?} 

Toward this end, we developed a novel methodology embodied in an end-to-end username \textbf{SQUA}tting \textbf{D}etection framework (\frameworkName{}) to facilitate the detection of potentially malicious confusion attempts against popular accounts. To do that, \frameworkName{} uses \systemName{} in combination with a classifier to output accounts that are confusing and require further analysis. 

\begin{figure*}[!htb]
\centering
\includegraphics[width=0.98\textwidth,height=\textheight,keepaspectratio]{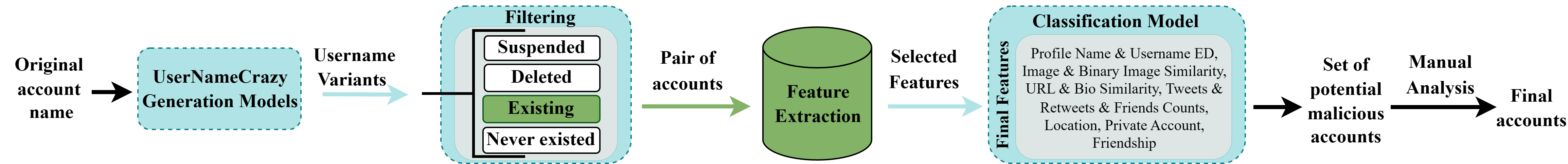}
\caption{High-level architecture of \frameworkName{}.}
\label{fig:end-to-end}
\vspace{-1em}
\end{figure*}

Figure~\ref{fig:end-to-end} illustrates the overall architecture of the framework. First, the seed account for which one wants to find confusing versions is selected. This is used as input in \systemName{} for the generation of squatted usernames (see Section~\ref{sec:measurement_study}). These are fed into a \textit{filtering} component which identifies active accounts corresponding to the squatted usernames~\footnote{The filtering also identifies suspended, deleted and non-existing accounts which might be useful for further analyses.}. 
The active accounts and the \textit{Initial Seed} account, are fed into a \textit{feature extraction} component which extracts certain account features. These are then passed into the \textit{classification} component. This outputs pairs of $<$\textit{initial, squatted}$>$ accounts where the squatted account is sufficiently confusing and requires further analysis. The username generation and filtering components were introduced during our measurement study. Next, we focus on the feature extraction and classification.


\subsection{Feature Extraction and Selection.}
\label{subsec:features}
\label{subsec:image_similarity_algorithm}
\vspace{5pt}\noindent\textbf{Initial Feature Set.} We start with an overview of the account features our classifier used. We manually select the features that could play a role in the account being confusing. This selection was based on our observations and findings from prior works identifying impersonation attempts and bot accounts~\cite{Minimum_weighted_feature, Consumption_Broadcast_bots, Botometer, Doppelganger_Bot_Attack}. 


We selected a number of features relevant to comparing the similarity of two accounts in terms of their image, bio, profile name, username and URL. For the \textit{Image Similarity Score}, we use an image recognition model, VGGFace2~\cite{VGGFace2}, to identify whether the squatted account's profile image is of the same person as the seed account's. It makes use of a convolutional neural network, known as SE-ResNet-50~\cite{SE-ResNet-50, VGGFace2}. 
To ensure accuracy of the model on our dataset we first performed a test on a manually labeled subset of $128$ images\footnote{We removed accounts that did not represent a human (e.g., companies and teams) or that portray groups of people, like bands. As a result, we are left with $64$ accounts. For each of these accounts we pick $1$ random generated variant and for every variant we select $2$ random profile pictures, resulting in $128$ profile pictures in total.\label{ref:64_accounts}}; this resulted in a $85.93\%$ test accuracy for the cutoff threshold $0.65$ (see Table~\ref{tab:VGG-Face2_table} in Appendix~\ref{appendix:vgg_face} for the full results). Note that all images with a higher score are classified as `\textit{no similarity}'. \ignore{Also t}The test accuracy is lower than previously reported ($90.8\%$)~\cite{VGGFace2} but still high enough to justify using this model for our prototype implementation. We also use \textit{Binary Image Similarity} which \ignore{simply }returns 0 and 1 if the similarity score is below or above the selected threshold respectively.

For \textit{bio similarity} we use the \textit{Jaccard Distance~\cite{Jaccard}} which is a similarity metric between two sets. The similarity index ranges from $0$ to $1$. The closer to $1$, the more similar the bios. Even though \textit{Jaccard similarity} does not have a notion of semantics, we found it to perform comparably with calculating the distance between the embeddings of bio statements, where we calculated $200$-dimensional embeddings~\cite{Glove_embeddings} pre-trained on Tweets. We also convert any special character (e.g., emojis) in to word format. For profile and username similarities we compute the Levenshtein distance~\cite{Levenshtein_distance}. To calculate URL similarity we collect the textual representation of a profile's ‘website' field for both \ignore{accounts}users. We first \ignore{examine}check whether the URLs are exactly the same or the squatted account uses a URL that includes the seed \ignore{user}name as a substring of it. If any of these is true, the result is ‘$1$'. 
If the \ignore{accounts}users use completely different URLs the result is ‘$0$'. \ignore{If only one of the 'website' fields are empty, the result is ‘$0.5$'.}

We also used several count-based features such as \textit{Friends Count}, \textit{Followers Count}, \textit{Tweet Count}, and \textit{Retweet Count}, and the features \textit{Private}, \textit{Friendship}, \textit{Location}, and \textit{Tweet Sentiment.} \textit{Private}, indicates whether an account is private. The \textit{Friendship} is a binary feature which shows if a squatted account follows the seed account. In terms of \textit{Location}, for every account, we collect the textual representation of the mentioned location, like the city name. We then check whether a squatted account shares the same location with the seed one. If yes, the returned results is ‘$1$' (this is applicable also if both locations are empty). Otherwise, the result is ‘$0$'. Lastly, for analyzing tweet sentiment we fetch the $500$ most recent tweets of the squatted account. We use NLTK~\cite{NLTK} for preprocessing the text to remove the stop-words, user-mentions and punctuation marks, and then we apply TextBlob~\cite{Textblob}, a sentiment method for classifying the tweets as \textit{neutral}, \textit{positive} and \textit{negative}. For the accounts with no tweets, the score of every category is ‘$0$'. Since in almost $95\%$ of the cases the returned result was ‘neutral', we discard this feature. Note also that we normalize all the data between $0$ and $1$
based on the min-max normalization process~\cite{Min_Max_Normalization}.

\vspace{3pt}\noindent\textbf{Feature Selection.} The initial set of features showed success in relevant classification tasks but not all of them are necessarily important for our task. In fact, redundant features can decrease the performance of the classifier by introducing noise~\cite{Removing_Noise_1, Removing_Noise_2}.
To remove \replaced[]{such}{redundant} features we select the optimal feature set via the \textit{Recursive Feature Elimination} (RFE) with Cross-Validation process~\cite{scikit-learn, Recursive_Feature_Elimination}, optimizing for accuracy. Our final feature set includes the \replaced[]{\textit{Profile Name and Username Edit Distance}}{\textit{Profile Name Edit Distance}, \textit{Username Edit Distance}}, \textit{Image Similarity Score}, \textit{Binary Image Similarity}, \textit{Friendship}, \replaced[]{\textit{Friends and Tweet Count}}{ \textit{Friends Count}, \textit{Tweet Count}}, \replaced[]{\textit{Bio and URL Similarity}}{\textit{Bio Similarity}, \textit{URL Similarity}}, \textit{Location}, \textit{Retweets Count} and \textit{Private Account} features.


\subsection{Generating the Dataset.} \label{sec:datageneration}
For each $<$seed, squatted$>$ account pair, the extracted features will be used in a binary classification whose output is the label \textit{benign} or \textit{suspicious}, where suspicious refers to an account that can contribute to confusion. The goal of a successful classifier is to have a low false positive (a positive being a suspicious account) rate, so that when the tool is used in combination with manual analysis it allows to find malicious activity on the OSN as efficiently as possible. 

A main challenge in our framework's classification modeling is that there are no publicly available datasets of impersonation attempts which could use to train and evaluate candidate models. Existing ones, such as~\cite{Bots_Dataset, Fake_Dataset, Spammer_Dataset}, are either datasets that consist of bots, or fake accounts which are not trying to imitate a specific profile. Our definition of confusion requires a different dataset as it encompasses any account that shares a sufficient number of features with the seed account including bots, impersonations and other accounts involved in malicious activity. 
Therefore, we created our own dataset by manually labelling a subset of the data records we collected for our measurement study. We use the same $1378$ \ignore{accounts}users which two raters manually and independently label as \textit{benign} or \textit{suspicious} (see Section~\ref{subsec:bot_analysis})\ignore{as described in Section~\ref{subsec:bot_analysis}}. To fix the imbalance of our dataset, we produce more examples from the minority class by applying the Synthetic Minority Over-sampling Technique (SMOTE)~\cite{SMOTE}. We then use our labeled accounts to train and evaluate $7$ popular binary classification models which our prototype implementation of \frameworkName{} supports. In Section~\ref{sec:evaluation} we compare their performance.

%% file: sections/6_evaluation.tex
\section{Evaluation}
\label{sec:evaluation}

The goal of \frameworkName{} is to speed up the process of finding malicious activity on the network by identifying confusing accounts that require further manual analysis. It uses \systemName{} to avoid searching the entire social graph by \emph{reducing} the search domain. The filtering component helps identify from that username search domain which variants exist on the network. Here, we first determine \emph{the effectiveness and efficiency of \systemName{} in identifying squatted accounts and how it compares with previous string squatting tools} (Section~\ref{subsec:competency_of_tools}). 
Then we analyze \emph{the accuracy of \frameworkName{} in classifying confusion accounts} (Section~\ref{subsec:classifier_performance}).


\vspace{-0.8em}

\subsection{\systemName{} Performance.}
\label{subsec:competency_of_tools}

\vspace{0pt}\noindent\textbf{Effectiveness.} First we evaluate the ability of \systemName{} to produce \emph{relevant} username variants and compare it with prior tools which leverage squatting techniques for generating complex strings, namely \textit{URLCrazy}~\cite{URLCrazy} and \textit{AppCrazy}~\cite{Mobile_App_Squatting}. We use the $10$ most popular users of our ‘\textit{Initial Seed}’ as input to the three generation tools and then apply our filtering component. We report the number of username variants produced, the number of those that actually exist on the network, the number of produced username variants that are already suspended by the network and the number of the existing impersonation attempts found based on our manual analysis. Table~\ref{tab:comparison_of_tools} summarizes our results and a full break down per user can be found on Table~\ref{tab:comparison_of_tools_full} in the Appendix~\ref{appendix:gen_models}.

\begin{table*}[!htp]
\caption{Competency of \textit{URLCrazy}, \textit{AppCrazy} and \textit{\systemName{}}.}
\label{tab:comparison_of_tools}
\setlength{\tabcolsep}{1pt}
\centering
\resizebox{\textwidth-1em}{!}{\begin{tabular}{c|cccc|cccc|cccc}
\hline
\multicolumn{1}{c}{\multirow{6}{*}{\textbf{Usernames}}} & \multicolumn{4}{c}{\textbf{URLCrazy}} & \multicolumn{4}{c}{\textbf{AppCrazy}} & \multicolumn{4}{c
}{\textbf{UsernameCrazy}} \\ \hline 
 & \multicolumn{1}{c}{\begin{tabular}[c]{@{}c@{}}\# of \\Generated\\ Usernames\end{tabular}} & \multicolumn{1}{c}{\begin{tabular}[c]{@{}c@{}}\# of \\ Active\\ Squatted\\ Usernames\end{tabular}} & \multicolumn{1}{c}{\begin{tabular}[c]{@{}c@{}}\# of \\ Suspended \\ Squatted \\ Usernames\end{tabular}} & \begin{tabular}[c]{@{}c@{}}\# of \\ Active\\ Impersonators\end{tabular} & \multicolumn{1}{|c}{\begin{tabular}[c]{@{}c@{}}\# of \\Generated\\ Usernames\end{tabular}} & \multicolumn{1}{c}{\begin{tabular}[c]{@{}c@{}}\# of \\ Active\\ Squatted\\ Usernames\end{tabular}} & \multicolumn{1}{c}{\begin{tabular}[c]{@{}c@{}}\# of \\ Suspended \\ Squatted \\ Usernames\end{tabular}} & \begin{tabular}[c]{@{}c@{}}\# of \\ Active\\ Impersonators\end{tabular} & \multicolumn{1}{|c}{\begin{tabular}[c]{@{}c@{}}\# of \\Generated\\ Usernames\end{tabular}} & \multicolumn{1}{c}{\begin{tabular}[c]{@{}c@{}}\# of \\ Active\\ Squatted\\ Usernames\end{tabular}} & \multicolumn{1}{c}{\begin{tabular}[c]{@{}c@{}}\# of \\ Suspended \\ Squatted \\ Usernames\end{tabular}} & \begin{tabular}[c]{@{}c@{}}\# of \\ Active\\ Impersonators\end{tabular} \\ \hline
\multicolumn{1}{c|}{\begin{tabular}[c]{@{}c@{}}\textbf{Total}\\ ($\textbf{10}$ \textbf{users})\end{tabular}} & \multicolumn{1}{c}{$\textbf{1,274}$} & \multicolumn{1}{c}{\begin{tabular}[c]{@{}c@{}}$\textbf{215}$\\ ($\textbf{16.87\%}$)\end{tabular}} & \multicolumn{1}{c}{\begin{tabular}[c]{@{}c@{}}$\textbf{49}$\\ ($\textbf{3.84\%}$)\end{tabular}} & $\textbf{10}$ & \multicolumn{1}{c}{$\textbf{436}$} & \multicolumn{1}{c}{\begin{tabular}[c]{@{}c@{}}$\textbf{293}$\\ ($\textbf{67.20\%}$)\end{tabular}} & \multicolumn{1}{c}{\begin{tabular}[c]{@{}c@{}}$\textbf{73}$\\ ($\textbf{16.74\%}$)\end{tabular}} & $\textbf{17}$ & \multicolumn{1}{c}{$\textbf{96,968}$} & \multicolumn{1}{c}{\begin{tabular}[c]{@{}c@{}}$\textbf{6,629}$\\ ($\textbf{6.83\%}$)\end{tabular}} & \multicolumn{1}{c}{\begin{tabular}[c]{@{}c@{}}$\textbf{1,901}$\\ ($\textbf{1.96\%}$)\end{tabular}} & $\textbf{88}$ \\ \hline

\end{tabular}}
\vspace{-7pt}
\end{table*}

We observe that \systemName{} exhibits better username generation coverage compared to existing methods. It generated $O(10^4)$ usernames compared to $O(10^3)$ and $O(10^2)$ from \textit{URLCrazy} and \textit{AppCrazy}. This is a result of \systemName{}'s enhancements with new models, \textit{model self-repetition}, and \textit{model stacking}. We also observe that compared to both prior methods \systemName{} identified one order of magnitude more variants that exist on the network and that are suspended. This showcases \systemName{}'s effectiveness in producing relevant string variations. There are several reasons for this. Firstly,\ignore{online social networks} OSNs have constraints on the number and type of characters a username can have. Prior tools do not take that into account and use generation methods (such as bit flipping or string rearrangement) which produce a large number of usernames that are invalid to begin with. Further, other tools produce strings that include a top-level domain either at the head or tail of the name making them again automatically incompatible. 

Note that \deleted[]{for \systemName{} }the ratio of existing or suspended usernames over all generated usernames is low. This is because a lot of the usernames are valid but have not been used yet. Nonetheless, it is important to be able to produce them to discover future instances of squatting username attempts.

\ignore{
\subsection{Competency of Existing Frameworks.}
\label{subsec:competency_of_Doppelgänger}
\sd{We could and probably should still mention without going into details why existing impersonation classifiers are ineffective or irrelevant.}
\tl{check now}

Closest to our work, is the framework of~\cite{Doppelganger_Bot_Attack}.
The authors found only $3$ impersonation attempts on celebrities whereas with \systemName{} we managed to identify a significantly larger number of such cases. Focusing on non-celebrity accounts, they propose a model that can detect whether a pair of accounts that portrays the same person is an impersonation attack or is owned by the same user. We briefly mention the differences in methodology: i) their framework requires the manual selection of similar profiles, ii) the different classification goal results in different features being used, iii) it is not applicable to celebrities as it is highly unlikely that a celebrity owns multiple accounts. Having said that, their classifier is ineffective comparing to ours because: i) they assume that when a fake account interacts in some way with the original account, this account should be managed by the same entity but for popular accounts this assumption does not need to hold, ii) they implicitly assume that fake accounts are close to each other in the network whereas in our case the social neighborhood overlap between the original and squatted account is irrelevant as followers of celebrity impersonations will not likely be overlapping with the followers of the original account and iii) we employ a more advanced image similarity technique which allows to recognise the same person in different environments and poses comparing to their generic matching techniques.
}

\begin{figure}[!t]
\centering
\includegraphics[width=\columnwidth,height=\textheight,keepaspectratio]{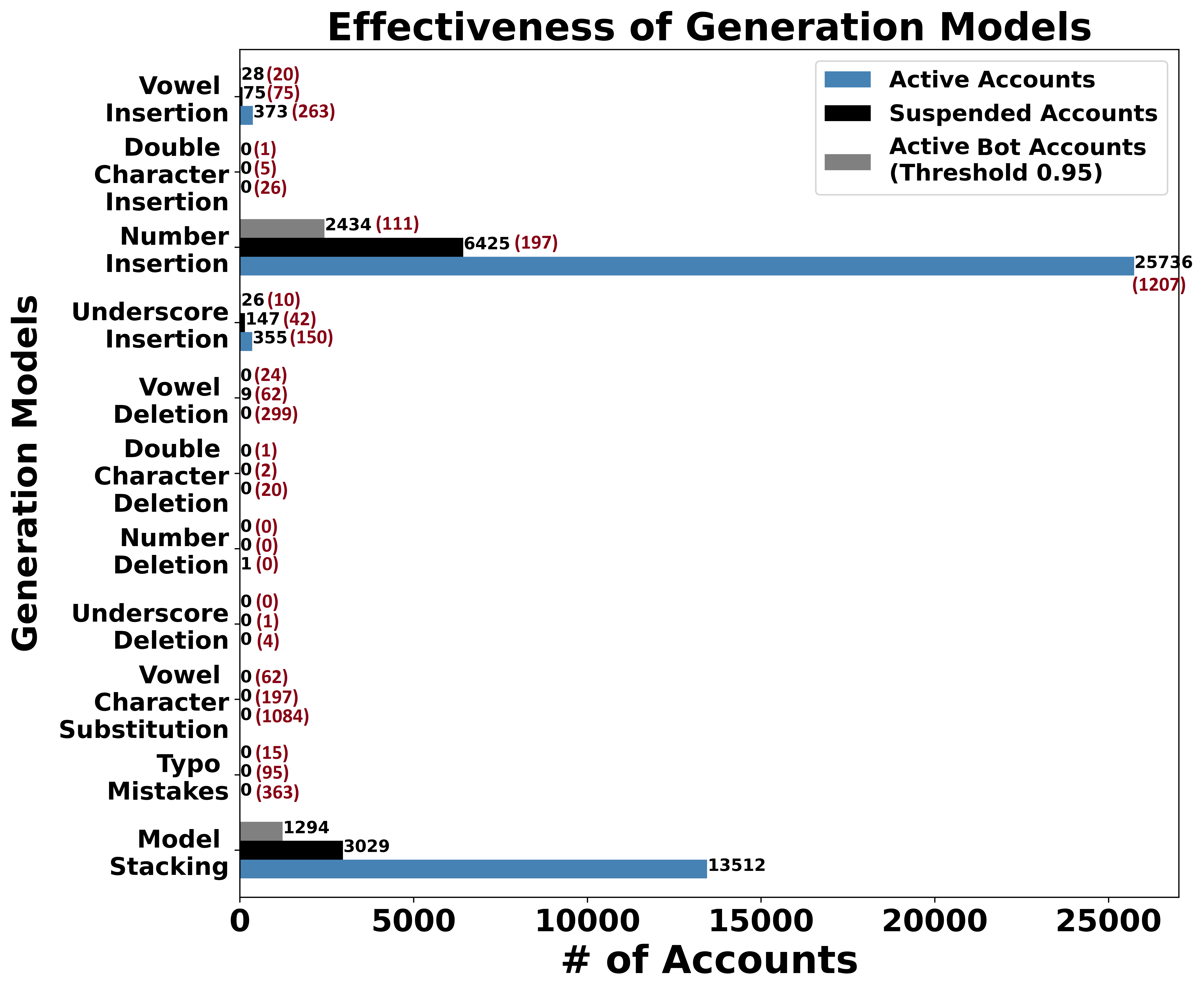}
\caption{The effectiveness of a) each generation model with \emph{self-repetition} and b) \emph{model stacking}. In \emph{parenthesis} is the effectiveness of the primitive models---not shown on the bars.}
\label{fig:Generation_models_efficiency}
\vspace{-5pt}
\end{figure}

\vspace{3pt}\noindent\textbf{Effectiveness of Generation Models. } \label{subsec:model_effectiveness}
Next, we want to better understand which of \systemName{}'s generation models are the most important. We repeat our effectiveness analysis for the \emph{primitive models}, the generation models with \emph{self-repetition} and the \emph{model stacking} of \systemName{} for all the usernames of the \emph{Initial Seed}. We compare the number of produced username variants that actually exist on the network or have already been suspended by the platform. \ignore{In addition, we}We also measure how many active \ignore{accounts}users found are marked as potential bots by Botometer~\cite{Botometer} as this is a further indication of the ability of \systemName{} to produce relevant usernames. We set Botometer's detection threshold to $0.95$ (see Section~\ref{subsec:bot_analysis}).  Figure~\ref{fig:Generation_models_efficiency} summarizes our results. We observe that a very large number of the generated variants have been either already suspended or declared as potential bots from Botometer. This demonstrates that our username generation models can reveal numerous squatted accounts that are potentially malicious. Moreover, we observe that the number insertion method is the most successful in identifying active accounts and that the majority of such accounts are either suspended or marked as bots. This reveals that a common username squatting tactic on OSNs is to add a digit as a prefix or as a suffix on a popular account's username. This strategy is simple to both conceptualize and implement. Considering also our measurements on edit distance (Section~\ref{subsec:characteristics}) we\ignore{can} observe that adding between $1$--$3$ digits on a target username seems to be the most common strategy.

\vspace{3pt}\noindent\textbf{\systemName{} Runtime Efficiency. }
Since \systemName{} can generate thousands of candidate usernames, it is important to do that efficiently. To measure \systemName{}'s runtime performance, we provided as input to \systemName{} the \emph{Initial Seed} (i.e., the usernames of the $97$ most popular\ignore{ accounts} users for \replaced[]{\emph{X}}{Twitter}). We set the maximum character count of a variant to be generated to be $15$ (which is also the maximum allowed for \replaced[]{\emph{X}}{Twitter} usernames)~\cite{Username_policy}. \systemName{} ran on commodity hardware (a laptop with an Intel Core i$5$-$8265$U processor\deleted{ at $1.6$ GHz} with $8$ GB of RAM). \systemName{} \replaced[]{produced}{was able to produce} $851,682$ valid usernames in $7.6$ seconds, which verifies that it can be a practical component in an end-to-end discovery of squatted accounts.



\subsection{Classification Performance.}
\label{subsec:classifier_performance}
\vspace{0pt}\noindent\textbf{Training and Evaluation Metrics.} \systemName{} produces relevant usernames but not all of them are used with malicious intent. Our classification component aims to take pairs of $<$seed, squatted$>$ accounts and decide whether the squatted version is suspicious. To evaluate the classification performance of the model we split our dataset to $70\%$ ($965$ users) and $30\%$ ($413$ users) for training and testing, respectively. Then, we evaluate $7$ of the most popular classification algorithms that have been shown to perform well on binary classification tasks: Random Forest~\cite{Random_Forest}, Naive Bayes~\cite{Naive_Bayes}, Logistic Regression~\cite{LR}, K-Nearest Neighbor (KNN)~\cite{KNN}, Support Machine Vector (SVM)~\cite{SVM}, Decision Tree~\cite{Decision_Tree}, Neural Network (NN)~\cite{Neural_Network}; all using the default parameters. 
\textit{Precision}, \textit{Recall} and \textit{F1-Score}~\cite{Precision_Recall_F1_Score} have been used as evaluation metrics. We found that Random Forest performed the best in all metrics. To improve its performance we fine tune its parameters by applying ‘\textit{Hyperparameter Tuning}~\cite{Hyperparameter_Tuning}' on all parameters, namely ‘\textit{n\_estimators, min\_samples\_split, min\_samples\_leaf, max\_features, max\_depth} and \textit{bootstrap}'. 

\vspace{3pt}\noindent\textbf{Overall Performance.} Random Forest performs better than the other classifiers, with $94\%$ and $94.5\%$ average precision and recall, respectively. Table~\ref{tab:classification_report} in Appendix~\ref{appendix:squad_performance} summarizes the classification performance results of all models. While not shown in the table, the accuracy of the Random Forest is $94.44\%$. The model correctly classifies $154/162$ \ignore{formula: (test size of malicious - false negative) / test size of malicious} accounts as malicious. 
We also compute the ‘\textit{Mean AUC}' score for all classification models and illustrate the trade-off between the false positive and true positive rates in Figure~\ref{fig:roc_curve} in Appendix~\ref{appendix:squad_performance}. We observe that there is an $99\%$ chance that Random Forest will correctly distinguish negative and positive classes, outperforming the other models. 
The number of false positives plays a crucial role in eliminating the manual analysis effort for identifying malicious attempts. After a closer inspection we find that our model returns $8$ false negative cases, but even though it misses a few cases we already demonstrated that our approach is vastly superior to existing squatting generation techniques while it is more practical than expensive deep graph traversals. We also find $15$ false positives which shows that \frameworkName{} keeps the number of false alarms low.

\vspace{3pt}\noindent\textbf{Improving \frameworkName{}.} \frameworkName{}'s dataset was limited to a relatively small number of manually labelled users. One potential improvement could be to increase the size of the labeled dataset. 

We also investigated the false positive (FP) and false negative (FN) cases. For the $15$ benign users \frameworkName{} misclassified (FP) we observed that a) all \ignore{accounts}users have small \emph{username} edit distance ($\leq3$), b) $7/15$ accounts have small \emph{profile name} edit distance ($\leq3$), c) $9/15$ profiles have similar image with the target and VGGFace$2$ returns a score lower than the threshold (marked as similar). These explain why \frameworkName{} predicted them as suspicious. Looking closer at the latter set, we observed that (d) $6/9$ use the word ‘fan' or ‘parody' in their \ignore{account }bio and $4/6$ of these exhibited a high bio similarity ($>$ $0.6$).
We leveraged these and introduced a post-filter on \frameworkName{} which uses a keyword search for ‘fan’\ignore{ and }/‘parody’ in account bios and negates the suspicious label if it finds a match. This reduced our false positives by almost $30\%$. We leave integrating more complex \ignore{techniques}methods based on word/sentence embeddings on the bios for future work.

For the $8$ malicious accounts \frameworkName{} missed (FN) we made the following observations: a) all accounts have small \emph{username} edit distance ($\leq3$), b) half of the accounts have small \emph{profile name} edit distance ($\leq3$), c) $6/8$ accounts have similar image with the target but VGGFace$2$ returns a score greater than the threshold and d) $4$ of the latter accounts have also very low bio similarity ($<$ $0.10$). As a result, the combination of the misclassification of the image and the low bio similarity lead to a wrong classification. Thus, \frameworkName{} can improve as VGGFace$2$ or image recognition improves---its modular design allows for simply exchanging the face recognition module with another one. \added[]{Also, \frameworkName{} only supports image similarity of faces and avatars. Hence, it can miss cases where an impersonator uses a different subject than a face in its profile photo, but as shown in prior work such strategies are less effective~\cite{profile_photos_1, profile_photos_2}.} Lastly, \frameworkName{} can \deleted[]{also }benefit from integrating more advanced models for characterizing users' bios, activity and tweets.

%% file: sections/7_applying_squad.tex
\section{Applying \frameworkName{}}
\label{sec:discussion}


\vspace{3pt}\noindent\textbf{Applying \frameworkName{} to Non-Popular Accounts.}\label{subsec:squad_non_popular} Using the CrowdFlower AI gender predictor dataset~\cite{kaggle_dataset_twitter_2} we randomly select $15$ non-private profiles with less than $200$ followers ($98\%$ of \replaced[]{\emph{X}}{Twitter} users have less than $400$ followers~\cite{inside_twitter}). Users without\ignore{ an avatar or} a face in their profile picture are disregarded from the selection. \frameworkName{} returns $159$ active accounts where $61$ have a face in their\ignore{profile} picture. Out of the latter, $0$ $(0\%)$ classified as \emph{suspicious} and $61$ $(100\%)$ as \emph{benign} accounts by \frameworkName{}. After manual verification, we find only $1$ \emph{false negative} case. The results contradict the insights of~\cite{Doppelganger_Bot_Attack}, where any account on \replaced[]{\emph{X}}{Twitter} can be a victim of impersonation. We summarize our findings in Table~\ref{tab:non_popular_accounts} (see Appendix~\ref{appendix:non_popular_accounts}).


\vspace{3pt}\noindent\textbf{Applying \frameworkName{} to Celebrities.}\label{subsec:squad_popular} We apply \frameworkName{} to $21$~\footref{ref:64_accounts} users of our \emph{Initial Seed} (see Section~\ref{subsec:data_collection}). \frameworkName{} returns $3,295$ active accounts where $1,137$ have a face in their profile picture. Out of the latter, $625$ $(55\%)$ classified as \emph{suspicious} and $512$ $(45\%)$ as \emph{benign} accounts by \frameworkName{}. Alarmingly, $297$ $(47.5\%)$\ignore{variants} users from the suspicious set were generated by the \emph{number insertion} strategy, signifying the model's malicious squatting intent (see Section~\ref{subsec:model_effectiveness}).

\vspace{3pt}\noindent\textbf{Analysing the Behavior of Suspicious Profiles.}\label{subsec:behaviour_of_impersonators} To \replaced[]{find if}{identify whether} the suspicious variants are engaged in malicious activities, we analysed the content of their tweets. Using the full-archive search endpoint (API v2~\cite{Twitter_Full_Archive}, see Section~\ref{subsec:user_mentions}) we fetch their $500$ most recent actual tweets, resulting in $8,516$ tweets where $1,476$ $(17.3\%)$ include a URL and $5,056$ $(59.3\%)$ are \emph{English} tweets. We manually observed two trends in the tweets of malicious \replaced[]{users}{accounts} which we further measured. Firstly, they share URLs which sometimes look suspicious and secondly they try to grow their follower base. In terms of URLs in tweets, we first analyze their integrity. Since none of the URLs are present in the VirusTotal database~\cite{virus_total} we submit them for scanning. Alarmingly, we discover: i) $24$ $(1.63\%)$ URLs were marked as malicious by $1$-$3$ antivirus systems, ii) $533$ $(36.1\%)$ URLs fail to use \emph{HTTPS}, and iii) $16$ $(1.08\%)$ URLs automatically obtained a TLS certificate from \emph{Let's Encrypt} Certificate Authority~\cite{letsencrypt}, a service that has\deleted[]{recently} faced criticism for its decision to abstain from implementing any form of security checks prior to granting certificates to domain owners~\cite{letsencrypt_problems}. We also measured follower base growth attempts in the English tweets using a simple keyword search approach, looking for the phrase \emph{follow me} (with slight variations acceptable). We found $266$ $(5.3\%)$ such cases, being posted by $78$ distinct users.

\vspace{3pt}\noindent\textbf{Username Squatting across Platforms.}\label{subsec:username_squatting_across_platforms} Username squatting can be exploited on other OSNs that use usernames as unique identifiers to search for and interact with users, such as Instagram and TikTok. To \replaced[]{illustrate}{demonstrate} the extent of the problem on other platforms we would ideally have to (a) adapt \systemName{} generation models to the username constraints of each platform, (b) collect existing variants on each platform and, (c) classify each account. The former is easy to do and can be efficiently executed. (b) on the other hand is harder especially since there are no publicly available APIs to allow us to automatically crawl Instagram and TikTok. Hence, this would require developing specialized web crawlers for each case. Lastly, (c) is also hard as it would entail manual analysis on either all or a relatively large subset of the existing variants. Instead, we design a more targeted experiment which can quickly allow us to collect preliminary indicators of the extent of the problem on other OSNs.\deleted[]{ We find that almost half of the squatted profiles that impersonate on \replaced[]{\emph{X}}{Twitter} exist on other social networks by sharing common properties.}\ignore{ We describe the measurement methodology and summarise our findings in Table~\ref{tab:impersonation_across_OSNs} in Appendix~\ref{appendix:username_squatting_on_other_platforms}.} 

\deleted[]{Furthermore, }We observe that \deleted[]{a lot of }popular profiles maintain accounts on other \replaced[]{OSNs}{platforms} with the same username. \replaced[]{Hence,}{Therefore,} if other \replaced[]{OSNs}{platforms} are plagued with similar issues, then it is likely that variants that were marked as malicious on \replaced[]{\emph{X}}{Twitter} could appear on \replaced[]{these OSNs}{the other platforms} too. To examine this, we select the top $10$ accounts of our seed and \added[]{manually} check how many exist with the same username across \deleted[]{both }Instagram and TikTok; this resulted in $5$ profiles. For each profile we use the squatted usernames, previously \replaced[]{generated by \systemName{}, }{found to belong to impersonator accounts (Section~\ref{subsec:bot_analysis}) to}\added[]{and find how many still exist on \emph{X}. Then, for every active variant we manually select the ones that are likely impersonators and} examine whether they a) exist with the same username, b) have a similar image and c) perform potential malicious activity on other OSNs. We call ‘\emph{Level $\textit{1}$}' the users who cover only the first property, ‘\emph{Level $\textit{2}$}' the profiles which cover the first property plus any of the other two and ‘\emph{Level $\textit{3}$}' the accounts that cover all the aforementioned properties.

Table~\ref{tab:impersonation_across_OSNs} summarises our findings.
\replaced[]{T}{We find that t}he most common pattern of the active squatted accounts across\deleted[]{ all} OSNs remains the \emph{Number Insertion} pattern. Lastly\ignore{More importantly},\deleted[]{ we observe that} almost half of the squatted accounts that impersonate on \replaced[]{\emph{X}}{Twitter} exist on other OSNs and have common properties. 

\begin{table}[H]
\caption{Similar impersonation attempts across different Social Networks.}
\label{tab:impersonation_across_OSNs}
\resizebox{\columnwidth}{!}{%
\begin{tabular}{lcccc}
\hline
\multicolumn{5}{c}{\textbf{Impersonations Across OSNs}} \\ \hline
\multicolumn{1}{c}{\textbf{Users}} &
  \multicolumn{1}{c}{\textbf{\begin{tabular}[c]{@{}c@{}}\replaced[]{\emph{X}}{Twitter} - Active\\ Impersonations\end{tabular}}} &
  \multicolumn{1}{c}{\textbf{Category}} &
  \multicolumn{1}{c}{\textbf{Instagram}} &
  \textbf{TikTok} \\ \hline
\multicolumn{1}{l}{\multirow{2}{*}{@katyperry}} &
  \multicolumn{1}{c}{\multirow{2}{*}{$7$}} &
  \multicolumn{1}{c}{Level $3$} &
  \multicolumn{1}{c}{$1/7$} &
  $0/7$ \\ 
\multicolumn{1}{l}{} &
  \multicolumn{1}{c}{} &
  \multicolumn{1}{c}{Level $2$} &
  \multicolumn{1}{c}{$4/7$} &
  $4/7$ \\ \cline{3-5} 
\multicolumn{1}{l}{\multirow{2}{*}{@justinbieber}} &
  \multicolumn{1}{c}{\multirow{2}{*}{$20$}} &
  \multicolumn{1}{c}{Level $3$} &
  \multicolumn{1}{c}{$3/20$} &
  $4/20$ \\ 
\multicolumn{1}{l}{} &
  \multicolumn{1}{c}{} &
  \multicolumn{1}{c}{Level $2$} &
  \multicolumn{1}{c}{$7/20$} &
  $10/20$ \\ \cline{3-5} 
\multicolumn{1}{l}{\multirow{2}{*}{@cristiano}} &
  \multicolumn{1}{c}{\multirow{2}{*}{$1$}} &
  \multicolumn{1}{c}{Level $3$} &
  \multicolumn{1}{c}{$0/1$} &
  $0/1$ \\ 
\multicolumn{1}{l}{} &
  \multicolumn{1}{c}{} &
  \multicolumn{1}{c}{Level $2$} &
  \multicolumn{1}{c}{$1/1$} &
  $1/1$ \\\cline{3-5} 
\multicolumn{1}{l}{\multirow{2}{*}{@ladygaga}} &
  \multicolumn{1}{c}{\multirow{2}{*}{$10$}} &
  \multicolumn{1}{c}{Level $3$} &
  \multicolumn{1}{c}{$0/10$} &
  $1/10$ \\ 
\multicolumn{1}{l}{} &
  \multicolumn{1}{c}{} &
  \multicolumn{1}{c}{Level $2$} &
  \multicolumn{1}{c}{$3/10$} &
  $7/10$ \\ \cline{3-5} 
\multicolumn{1}{l}{\multirow{2}{*}{@kimkardashian}} &
  \multicolumn{1}{c}{\multirow{2}{*}{$5$}} &
  \multicolumn{1}{c}{Level $3$} &
  \multicolumn{1}{c}{$2/5$} &
  $0/5$ \\ 
\multicolumn{1}{l}{} &
  \multicolumn{1}{c}{} &
  \multicolumn{1}{c}{Level $2$} &
  \multicolumn{1}{c}{$3/5$} &
  $0/5$ \\ \cline{3-5} 
\multicolumn{1}{c}{\multirow{3}{*}{\textbf{Total}}} &
  \multicolumn{1}{c}{\multirow{3}{*}{\textbf{$\textbf{42}$}}} &
  \multicolumn{1}{c}{\textbf{Level $\textbf{3}$}} &
  \multicolumn{1}{c}{\textbf{$\textbf{6/42}$}} &
  \textbf{$\textbf{5/42}$} \\ 
\multicolumn{1}{c}{} &
  \multicolumn{1}{c}{} &
  \multicolumn{1}{c}{\textbf{Level $\textbf{2}$}} &
  \multicolumn{1}{c}{\textbf{$\textbf{18/42}$}} &
  \textbf{$\textbf{22/42}$} \\ 
\multicolumn{1}{c}{} &
  \multicolumn{1}{c}{} &
  \multicolumn{1}{c}{\textbf{Level $\textbf{1}$}} &
  \multicolumn{1}{c}{\textbf{$\textbf{27/42}$}} &
  \textbf{$\textbf{28/42}$} \\ \hline
\end{tabular}}
\vspace{-10pt}
\end{table}

%% file: sections/8_background_work.tex
\section{Related Work}
\label{sec:background}

\vspace{0pt}\noindent\textbf{Fake Accounts on OSNs.} Prior work cloned OSN profiles without proposing ways to detect them~\cite{Automated_Identity_Theft} and studied the prevalence of profile name reuse on OSNs~\cite{Profile_Name_Reuse,Profile_Name_Reuse_2}. Researchers have presented ways for detecting Sybil attacks using e.g., visual profile similarity~\cite{New_Approach_Profile_Cloning}, string matches and users' relationship measurements to identify whether one's account has been a victim of identity theft~\cite{Kontaxis_Social_Networks}, the network relationship and attribute similarity~\cite{Active_Detection_Of_Identity_Theft}\replaced[]{,}{ and} honest networks to reveal dishonest nodes via an inference engine~\cite{Danezis2009SybilInferDS}\deleted[]{ and graph properties to rank profiles\ignore{according to} based on their perceived likelihood of being fake~\cite{Aiding_the_Detection_of_Fake_Accounts}}.
\ignore{.~\cite{Comprehensive_study_of_sybil_Attacks} presented a comprehensive comparison between proposed Sybil defenses until $2011$.} In contrast, \frameworkName{} combines username squatting techniques and static profile features for malicious classification.

\vspace{3pt}\noindent\textbf{Impersonation on OSNs.} 
\label{subsec:impersonation_background}
Several works analyzed posts~\cite{Impersonation_on_Social_Media} and post reactions~\cite{Deep_Dive_on_Impersonating} to identify impersonation of celebrities on Instagram.
Closest to our work, Goga et al.~\cite{Doppelganger_Bot_Attack} detect whether a pair of accounts on \replaced[]{\emph{X}}{Twitter} that portray the same person is an impersonation attack or whether one identity owns both accounts (avata-avatar pair).
\replaced[]{T}{However, t}he proposed technique requires manual selection of similar profiles and relies on expensive social graph traversals. It also found that impersonation is not a prevalent threat for celebrities. We show the opposite. Notably, we are the first to demonstrate that username squatting is a central strategy to such attacks and use it to design an efficient detection tool. Also, Goga et al. argue that an avatar-avatar pair is not considered dangerous. However, most of our malicious cases (see Table~\ref{tab:malicious_users_raters}) belong to \replaced[]{this}{the above} category ($521$/$540$ profiles) making these users an overlooked problem. Finally, \frameworkName{} does not rely on the following assumptions: a) an impersonator should have a connection with its target (e.g., \deleted[]{being }friends, commenting\deleted[]{/reacting} on tweets) and b) impersonators are in the neighborhood of impersonators. \frameworkName{} can detect impersonators anywhere in the social graph.

\vspace{3pt}\noindent\textbf{Spam and Abusive Accounts on OSNs.}
Zheng et al.~\cite{Detecting_Spammers} create passive honey-profiles, log the friend requests/messages and find anomalous behavior. \added[]{Yang et al.}~\cite{Die_free_or_live_hard} use honeypots\ignore{ and blacklists} to identify spammers, extract common evasion tactics based on their behavior, evaluate proposed detection techniques~\cite{Evasion_tactic_1, Evasion_tactic_2, Evasion_tactic_3} and develop a system to detect spammers \replaced[]{on}{in} \replaced[]{\emph{X}}{Twitter}. \added[]{Others}~\cite{@spam,Warningbird,Design_Evaluation_of_a_Real-Time_URL_Spam} focus on developing suspicious URL detection systems\ignore{ for Twitter and other web services}. \added[]{Xu et al.}~\cite{Deep_entity_classification} propose a heavyweight ML framework which leverages features based on the\ignore{ social graph and the} direct/indirect neighbor properties to find users who violate the policies of an OSN, including bots\replaced[]{/spammers/fake users}{, spammers and fake users}. In contrast, the features used in \frameworkName{} \replaced[]{are}{were chosen to be} easy and computationally cheap to collect and were based on analyzing specifically\ignore{ impersonation accounts} impersonators.

\vspace{3pt}\noindent\textbf{Bot Accounts on OSNs.}
\ignore{Previous studies on identifying and profiling bots}\added[]{Previous studies}~\cite{Amato2018RecognizingHB, Temporal_Patterns_in_Bot_Activities, Frustrate_Twitter_from_automation, Detecting_and_Analyzing_Automated_Activity_on_Twitter, Combating_the_evolving_spammers} have focused on the behavior patterns of bot accounts, concluding that they confuse users and pose social, financial and political hazards\ignore{~\cite{Amato2018RecognizingHB, Temporal_Patterns_in_Bot_Activities, Frustrate_Twitter_from_automation, Detecting_and_Analyzing_Automated_Activity_on_Twitter, Combating_the_evolving_spammers}}. \added[]{Besides, Oentaryo et al.}~\cite{Consumption_Broadcast_bots} focus on benign bots; tools also exist for \deleted[]{profiling malicious bots~\cite{Botometer} and }detecting\ignore{ potential} misinformation\ignore{spreading}~\cite{BotSlayer}. In our work we use Botometer~\cite{Botometer} to classify existing squatted accounts and\ignore{ gain a better understanding of} understand their behavior. However, Botometer uses the characteristics of bots~\cite{Twitter_Botometer_Critique}, and while \textit{botness} is a good signal it is not sufficient for detecting impersonators\ignore{neither necessary nor sufficient for detecting the more general confusing accounts\ignore{ as well as} and impersonations}. Instead, we believe that the above works deviate from the content of our study\ignore{, which focuses on\ignore{ impersonation attempts} impersonators while treating any bot only as a \emph{suspicious} account}, rendering direct comparisons inappropriate.  

\vspace{3pt}\noindent\textbf{Squatting Techniques.} Squatting issues have been studied in different domains~\cite{Typo_Squatting, Sound_Squatting, Combo_Squatting, Mobile_App_Squatting, Skill_Squatting_Alexa, Bitsquatting_Phenomenon, zhang2019dangerous, Email_Typosquatting}, exploring how they trick users into illegal activities\ignore{ and have been examined on how can deceive\ignore{ users and lead them} users to become victims of illegal activities}. \deleted[]{Particularly, }\added[]{Panagiotis et al.}~\cite{Combo_Squatting} focus on combosquatting\ignore{,a technique first studied by an industry whitepaper~\cite{White_paper_combo_Squatting},} which\ignore{ in contrast to typosquatting it} lacks a generative model. An adversary can postfix\ignore{ and/or}/prefix any word constrained solely by the character limits\ignore{ maximum number of allowed characters}~\cite{Username_policy}, resulting in an exponential increase in the number of users that need to be searched within \replaced[]{a}{the \replaced[]{\emph{X}}{Twitter}} graph. Similar to \deleted[]{our }\systemName{}, tools also exist for generating string variations given a seed string~\cite{URLCrazy, Mobile_App_Squatting, DNStwist} but they suffer from several limitations \ignore{However, these are limited in their ability to generate compatible OSN usernames}\ignore{as they suffer from several limitations}\deleted[]{(see Section~\ref{subsec:competency_of_tools})}. To the best of our knowledge we are the first to a) develop a string generation tool compatible with OSN usernames, b) apply squatting methods to discover and characterize squatted usernames on OSNs and c) study how adversaries can adapt them to create confusion.

%% file: sections/9_conclusion.tex
\section{Conclusion}
\label{sec:conclusion}
Our study is the first to characterize \textit{username squatting} on OSNs. We showed that \textit{username squatting} can be used for confusion and impersonation. We found hundreds of
thousands of \textit{typo-mentions} mentioning squatted versions of popular accounts with no apparent mentioner--mentionee relationship, and demonstrated how the search recommendation algorithm of a popular OSN can unwittingly amplify online confusion. We discovered that thousands of squatted usernames are already suspended by \replaced[]{\emph{X}}{Twitter} while tens of thousands of \emph{active} squatted accounts are likely bots. While not all username variants are malicious, we show that \textit{username squatting} can be a useful signal for identifying such accounts. We designed \frameworkName{}, to aid detection of malicious accounts on OSNs. \frameworkName{} incorporates new techniques for username squatting along with a\ignore{novel} similarity feature extractor and classification components. \frameworkName{} can generate tens of thousands of relevant squatted usernames of a given account in a few seconds, and can achieve $94\%$ F1-score in detecting suspicious squatted accounts when trained on a small dataset. \added[]{\frameworkName{} can be used by account owners to identify attempts to impersonate them, by regulating authorities such as the Federal Trade Commission (FTC, US)~\cite{Federal_Trade_Commission} or the Information Commissioner's Office (ICO, UK)~\cite{Information_Commissioner_Office}, and the OSN platforms themselves who want to take measures to limit the negative impact impersonation and confusion can have on the community.} 
Lastly, we revealed that username squatting targets primarily popular accounts and $36\%$ of the suspicious accounts' tweets share insecure URLs and $5\%$ of the tweets try to grow their or others' follower base.

%% file: sections/appendix.tex
\label{sec:appendix}


\appendix



\ignore{

\section{User Classification}
\label{apppendix:user_classification}
Table~\ref{tab:Append_Table 1} depicts the categorization of the \emph{Initial Seed} of Twitter users into account types, as determined by three independent raters following the official Twitter documentation.

\begin{table*}[h!]
\caption{User categorization based on Twitter's policy. The accounts ‘@realdonaldtrump', ‘@arianagrande' and ‘@aamir\_khan' shown in light grey are not included in our experiments.}
\resizebox{\textwidth}{!}{%
\begin{tabular}{|l|l|}
\hline
\multicolumn{2}{|c|}{\textbf{User Classification}}                                          \\ \hline

\textbf{\begin{tabular}[c]{@{}l@{}}Government:\\ (5 accounts)\end{tabular}} &
  \begin{tabular}[c]{@{}l@{}}@barackobama, @narendramodi, @pmoindia, @potus, @hillaryclinton, \textcolor{lightgray}{@realdonaldtrump}\end{tabular} \\ \hline

\textbf{\begin{tabular}[c]{@{}l@{}}Companies, brands,\\and organizations:\\ (5 accounts)\end{tabular}} & @youtube, @twitter, @instagram, @nasa, @google
 \\ \hline
\textbf{\begin{tabular}[c]{@{}l@{}}News organizations\\ and journalists:\\ (10 accounts)\end{tabular}} &
  \begin{tabular}[c]{@{}l@{}}@cnnbrk, @nytimes, @CNN, @bbcbreaking, @sportscenter, @espn, @bbcworld, @theeconomist,\\ @reuters, @natgeo\end{tabular} \\ \hline

\textbf{\begin{tabular}[c]{@{}l@{}}Entertainment:\\ (60 accounts)\end{tabular}} &
  \begin{tabular}[c]{@{}l@{}}@katyperry, @justinbieber, @rihanna, @taylorswift13, @ladygaga, @theellenshow, @blakeshelton,\\ @jtimberlake, @kimkardashian, @selenagomez, @britneyspears, @shakira, @jimmyfallon, @drake,\\ @mileycyrus, @jlo, @brunomars, @oprah, @iamsrk, @srbachchan,  @niallofficial, @ricky\_martin,\\ @beingsalmankhan, @kevinhart4real, @wizkhalifa, @liltunechi, @harry\_styles, @louis\_tomlinson,\\ @akshaykumar, @liampayne, @pink, @chrisbrown, @onedirection,  @aliciakeys, @kyliejenner,\\ @kanyewest, @emmawatson, @conanobrien, @kendalljenner, @zaynmalik, @khloekardashian, \\ @adele, @ihrithik, @actuallynph, @deepikapadukone, @pitbull, @danieltosh, @priyankachopra,\\ @kourtneykardash, @shawnmendes, @coldplay, @bts\_twt, @eminem, @arrahman, @nickiminaj,\\ @avrillavigne, @mariahcarey, @davidguetta, @anushkasharma, \textcolor{lightgray}{@arianagrande}, \textcolor{lightgray}{@aamir\_khan} \end{tabular} \\ \hline

\textbf{\begin{tabular}[c]{@{}l@{}}Sports and gaming:\\ (15 accounts)\end{tabular}} &
  \begin{tabular}[c]{@{}l@{}}@cristiano, @neymarjr, @kingjames, @realmadrid, @imvkohli, @fcbarcelona, @sachin\_rt, @kaka, \\ @manutd, @nba, @championsleague, @nfl, @mesutozil1088, @andresiniesta8, @premierleague \end{tabular} \\ \hline

\textbf{\begin{tabular}[c]{@{}l@{}}Activists, organizers,\\and other influential\\ individuals:\\(3 accounts)\end{tabular}} & @billgates, @elonmusk, @mohamadalarefe \\ \hline

\end{tabular}
}

\label{tab:Append_Table 1}
\end{table*}
}

\section{\systemName{}'s Generation Models}
\label{appendix:gen_models}

In Section~\ref{subsec:username_variant_generation} we introduced \systemName{}'s taxonomy of generation models. Here we briefly describe each model.


    \vspace{2pt}\noindent \textit{\textbf{Vowel Insertion}}: Inserts an extra and same vowel character when it finds one, e.g., ‘@AxlRose’ into ‘@AaxlRose’.
    
    \vspace{2pt}\noindent \textit{\textbf{Double Character Insertion}}: After finding two consecutive and identical characters, it inserts the same character next to them, e.g., ‘@CNNbrk’ into ‘@CNNNbrk’.
    
    \vspace{2pt}\noindent \textit{\textbf{Number Insertion}}: Adds a number (the same or a different one) both in the end and at the beginning of a username, ‘@Cristiano’ into ‘@Cristiano$21$’ or ‘@$9$Cristiano’.
    
    \vspace{2pt}\noindent \textit{\textbf{Underscore Insertion}}: Inserts an underscore both in the end and at the beginning of a username, e.g., ‘@NBA’ into ‘@NBA\_’ and ‘@\_NBA’.
    
    \vspace{2pt}\noindent \textit{\textbf{Vowel Deletion}}: Deletes a vowel character when it finds one, e.g.,  ‘@BarackObama’ into ‘@BrackObama’.
    
    \vspace{2pt}\noindent \textit{\textbf{Double Character Deletion}}: Deletes two consecutive and identical characters, e.g., ‘@Twitter’ into ‘@Twier’.
    
    \vspace{2pt}\noindent \textit{\textbf{Number Deletion}}: Deletes all the numbers of the username, one each time, starting both from the end and the beginning of a username, e.g., ‘@AndresIniesta$8$’ into ‘@AndresIniesta’.
    
    \vspace{2pt}\noindent \textit{\textbf{Underscore Deletion}}: Deletes an underscore when it finds one, e.g., ‘@Ricky\_Martin’ into ‘@RickyMartin’.
    
    \vspace{2pt}\noindent \textit{\textbf{Vowel Character Substitution}}: Replaces a vowel with one of the rest vowel characters each time, e.g., ‘@BarackObama’ into ‘@BerackObama’.
    
    \vspace{2pt}\noindent \textit{\textbf{Common Misspellings/Homoglyphs}}: Specific characters are being replaced according to various recognized misspelling patterns \cite{Letter_and_symbol_misrecognition}, e.g., ‘@BarackObama’ into ‘@BarakObama’. In this experiment, we used a small number of common mistakes to remain time-efficient.

\added[]{In Section~\ref{subsec:competency_of_tools} we analyze the effectiveness of \systemName{}. Table~\ref{tab:comparison_of_tools_full} shows the number of the generated, active, suspended and active impersonator variants for each of the $10$ most popular users.}

\begin{table*}[!htbp]
\caption{Competency of \textit{URLCrazy}, \textit{AppCrazy} and \textit{\systemName{}}.}
\label{tab:comparison_of_tools_full}
\setlength{\tabcolsep}{1pt}
\centering
\resizebox{\textwidth-1em}{!}{\begin{tabular}{l|cccccccccccc}
\hline
\multicolumn{1}{c}{\multirow{6}{*}{\textbf{Usernames}}} & \multicolumn{4}{c}{\textbf{URLCrazy}} & \multicolumn{4}{c}{\textbf{AppCrazy}} & \multicolumn{4}{c
}{\textbf{UsernameCrazy}} \\ \hline 
 & \multicolumn{1}{c}{\begin{tabular}[c]{@{}c@{}}\# of \\Generated\\ Usernames\end{tabular}} & \multicolumn{1}{c}{\begin{tabular}[c]{@{}c@{}}\# of \\ Active\\ Squatted\\ Usernames\end{tabular}} & \multicolumn{1}{c}{\begin{tabular}[c]{@{}c@{}}\# of \\ Suspended \\ Squatted \\ Usernames\end{tabular}} & \begin{tabular}[c]{@{}c@{}}\# of \\ Active\\ Impersonators\end{tabular} & \multicolumn{1}{|c}{\begin{tabular}[c]{@{}c@{}}\# of \\Generated\\ Usernames\end{tabular}} & \multicolumn{1}{c}{\begin{tabular}[c]{@{}c@{}}\# of \\ Active\\ Squatted\\ Usernames\end{tabular}} & \multicolumn{1}{c}{\begin{tabular}[c]{@{}c@{}}\# of \\ Suspended \\ Squatted \\ Usernames\end{tabular}} & \begin{tabular}[c]{@{}c@{}}\# of \\ Active\\ Impersonators\end{tabular} & \multicolumn{1}{|c}{\begin{tabular}[c]{@{}c@{}}\# of \\Generated\\ Usernames\end{tabular}} & \multicolumn{1}{c}{\begin{tabular}[c]{@{}c@{}}\# of \\ Active\\ Squatted\\ Usernames\end{tabular}} & \multicolumn{1}{c}{\begin{tabular}[c]{@{}c@{}}\# of \\ Suspended \\ Squatted \\ Usernames\end{tabular}} & \begin{tabular}[c]{@{}c@{}}\# of \\ Active\\ Impersonators\end{tabular} \\ \hline
@barackobama & \multicolumn{1}{c}{$135$} & \multicolumn{1}{c}{$20$} & \multicolumn{1}{c}{$6$} & $0$ & \multicolumn{1}{|c}{$57$} & \multicolumn{1}{c}{$35$} & \multicolumn{1}{c}{$12$} & $0$ & \multicolumn{1}{|c}{$9,800$} & \multicolumn{1}{c}{$307$} & \multicolumn{1}{c}{$104$} & $1$ \\ 
@katyperry & \multicolumn{1}{c}{$118$} & \multicolumn{1}{c}{$17$} & \multicolumn{1}{c}{$9$} & $0$ & \multicolumn{1}{|c}{$25$} & \multicolumn{1}{c}{$15$} & \multicolumn{1}{c}{$5$} & $0$ & \multicolumn{1}{|c}{$5,992$} & \multicolumn{1}{c}{$640$} & \multicolumn{1}{c}{$500$} & $7$ \\ 
@justinbieber & \multicolumn{1}{c}{$170$} & \multicolumn{1}{c}{$34$} & \multicolumn{1}{c}{$4$} & $3$ & \multicolumn{1}{|c}{$60$} & \multicolumn{1}{c}{$52$} & \multicolumn{1}{c}{$7$} & $5$ & \multicolumn{1}{|c}{$9,010$} & \multicolumn{1}{c}{$555$} & \multicolumn{1}{c}{$110$} & $20$ \\ 
@rihanna & \multicolumn{1}{c}{$94$} & \multicolumn{1}{c}{$21$} & \multicolumn{1}{c}{$3$} & $0$ & \multicolumn{1}{|c}{$38$} & \multicolumn{1}{c}{$32$} & \multicolumn{1}{c}{$2$} & $1$ & \multicolumn{1}{|c}{$13,086$} & \multicolumn{1}{c}{$1,113$} & \multicolumn{1}{c}{$199$} & $9$ \\ 
@taylorswift13 & \multicolumn{1}{c}{$172$} & \multicolumn{1}{c}{$22$} & \multicolumn{1}{c}{$9$} & $2$ & \multicolumn{1}{|c}{$36$} & \multicolumn{1}{c}{$25$} & \multicolumn{1}{c}{$8$} & $2$ & \multicolumn{1}{|c}{$5,766$} & \multicolumn{1}{c}{$102$} & \multicolumn{1}{c}{$23$} & $5$ \\ 
@cristiano & \multicolumn{1}{c}{$131$} & \multicolumn{1}{c}{$34$} & \multicolumn{1}{c}{$1$} & $0$ & \multicolumn{1}{|c}{$49$} & \multicolumn{1}{c}{$46$} & \multicolumn{1}{c}{$1$} & $0$ & \multicolumn{1}{|c}{$17,687$} & \multicolumn{1}{c}{$1,714$} & \multicolumn{1}{c}{$134$} & $1$ \\ 
@ladygaga & \multicolumn{1}{c}{$96$} & \multicolumn{1}{c}{$18$} & \multicolumn{1}{c}{$2$} & $0$ & \multicolumn{1}{|c}{$32$} & \multicolumn{1}{c}{$26$} & \multicolumn{1}{c}{$6$} & $0$ & \multicolumn{1}{|c}{$6,821$} & \multicolumn{1}{c}{$954$} & \multicolumn{1}{c}{$287$} & $10$ \\ 
@theellenshow & \multicolumn{1}{c}{$168$} & \multicolumn{1}{c}{$5$} & \multicolumn{1}{c}{$6$} & $1$ & \multicolumn{1}{|c}{$51$} & \multicolumn{1}{c}{$13$} & \multicolumn{1}{c}{$8$} & $2$ & \multicolumn{1}{|c}{$6,999$} & \multicolumn{1}{c}{$46$} & \multicolumn{1}{c}{$62$} & $7$ \\ 
@youtube & \multicolumn{1}{c}{$101$} & \multicolumn{1}{c}{$30$} & \multicolumn{1}{c}{$2$} & $3$ & \multicolumn{1}{|c}{$41$} & \multicolumn{1}{c}{$31$} & \multicolumn{1}{c}{$9$} & $5$ & \multicolumn{1}{|c}{$13,057$} & \multicolumn{1}{c}{$1,078$} & \multicolumn{1}{c}{$353$} & $12$ \\ 
@jtimberlake & \multicolumn{1}{c}{$156$} & \multicolumn{1}{c}{$15$} & \multicolumn{1}{c}{$8$} & $1$ & \multicolumn{1}{|c}{$47$} & \multicolumn{1}{c}{$18$} & \multicolumn{1}{c}{$15$} & $2$ & \multicolumn{1}{|c}{$8,750$} & \multicolumn{1}{c}{$120$} & \multicolumn{1}{c}{$129$} & $16$ \\ \hline
\multicolumn{1}{c}{\begin{tabular}[c]{@{}c@{}}\textbf{Total}\\ ($\textbf{10}$ \textbf{users})\end{tabular}} & \multicolumn{1}{c}{$\textbf{1,274}$} & \multicolumn{1}{c}{\begin{tabular}[c]{@{}c@{}}$\textbf{215}$\\ ($\textbf{16.87\%}$)\end{tabular}} & \multicolumn{1}{c}{\begin{tabular}[c]{@{}c@{}}$\textbf{49}$\\ ($\textbf{3.84\%}$)\end{tabular}} & $\textbf{10}$ & \multicolumn{1}{c}{$\textbf{436}$} & \multicolumn{1}{c}{\begin{tabular}[c]{@{}c@{}}$\textbf{293}$\\ ($\textbf{67.20\%}$)\end{tabular}} & \multicolumn{1}{c}{\begin{tabular}[c]{@{}c@{}}$\textbf{73}$\\ ($\textbf{16.74\%}$)\end{tabular}} & $\textbf{17}$ & \multicolumn{1}{c}{$\textbf{96,968}$} & \multicolumn{1}{c}{\begin{tabular}[c]{@{}c@{}}$\textbf{6,629}$\\ ($\textbf{6.83\%}$)\end{tabular}} & \multicolumn{1}{c}{\begin{tabular}[c]{@{}c@{}}$\textbf{1,901}$\\ ($\textbf{1.96\%}$)\end{tabular}} & $\textbf{88}$ \\ \hline

\end{tabular}}
\end{table*}


\ignore{

\section{Bot description}
\label{appendix:bot_types}
In Section~\ref{subsec:bot_analysis} we applied Botometer to all the active accounts to validate our hypothesis that a significant number of them are malicious/bots users. Table~\ref{tab:Botometer_types} shows the description of Botometer's bot account sub-categories. 

\begin{table}[H]
\caption{Description of Botometer's bot types~\cite{Botometer}}
\resizebox{\columnwidth}{!}{%
\begin{tabular}{cc}
\hline
\multicolumn{2}{c}{\textbf{Botometer - Bot Types}}                                                  \\ \hline
\multicolumn{1}{c}{\textbf{Types}} & \textbf{Description}                       \\ \hline
\multicolumn{1}{c}{Astroturf}     & Political bot account                      \\ 
\multicolumn{1}{c}{Fake follower} & Bots purchased to increase follower counts \\ 
\multicolumn{1}{c}{Financial}     & Bots that post using cash-tags             \\ 
\multicolumn{1}{c}{Self declared} & Bots from botwiki.org                      \\ 
\multicolumn{1}{c}{Spammer} & \begin{tabular}[c]{@{}c@{}}Accounts labeled as spambots from datasets\end{tabular}                            \\ 
\multicolumn{1}{c}{Other}   & \begin{tabular}[c]{@{}c@{}}Miscellaneous other bots obtained from manual \\ annotation, user feedback, etc.\end{tabular} \\ \hline
\end{tabular}
}
\label{tab:Botometer_types}
\end{table}

}

\section{VGGFace2 \& Thresholds}
\label{appendix:vgg_face}
In Section~\ref{subsec:image_similarity_algorithm} we described how we ensured the accuracy of VGGFace2 stays at a high enough score. Table~\ref{tab:VGG-Face2_table} depicts the accuracy results of VGGFace2 with different thresholds. \added[]{We observe that the highest test accuracy on our dataset was $85.93\%$ when the threshold was set to $0.65$.}

\begin{table}[!htbp] 
\caption{Accuracy of VGGFace2 with various thresholds. Setting the threshold value to 0.65, the model has the best accuracy.}
\resizebox{\columnwidth}{!}{%
\begin{tabular}{ccccc}
\hline
\multicolumn{5}{c}{\textbf{VGGFace2}}                                                                                 \\ \hline
\multicolumn{1}{l}{\textbf{Threshold}} &
  \multicolumn{1}{l}{\textbf{Accuracy}} &
  \multicolumn{1}{l}{\textbf{\# of Images}} &
  \multicolumn{1}{l}{\textbf{\# of FP}} &
  \multicolumn{1}{l}{\textbf{\# of FN}} \\ \hline
\multicolumn{1}{c}{\textbf{0.5}}  & \multicolumn{1}{c}{$82.81\%$} & \multicolumn{1}{c}{$128$} & \multicolumn{1}{c}{$9$} & $13$ \\ 
\multicolumn{1}{c}{\textbf{0.6}}  & \multicolumn{1}{c}{$84.37\%$} & \multicolumn{1}{c}{$128$} & \multicolumn{1}{c}{$8$} & $12$ \\ 
\multicolumn{1}{c}{\textbf{0.65}} & \multicolumn{1}{c}{$85.93\%$} & \multicolumn{1}{c}{$128$} & \multicolumn{1}{c}{$6$} & $12$  \\ 
\multicolumn{1}{c}{\textbf{0.7}}  & \multicolumn{1}{c}{$85.15\%$} & \multicolumn{1}{c}{$128$} & \multicolumn{1}{c}{$7$} & $12$ \\ \hline
\end{tabular}
}
\label{tab:VGG-Face2_table}
\end{table}

\ignore{
\section{More examples}
\label{appendix:more_examples}

In Section~\ref{sec:problem} we defined the username squatting problem in OSNs. We further present several examples of potential impersonation attempts we revealed. Figures~\ref{fig:David_Guetta_Impersonation},~\ref{fig:malicious_links_profile},~\ref{fig:theellenshow_impersonation},~\ref{fig:shakira_impersonation} demonstrate the profiles of impersonation attempts, Figure~\ref{fig:fake_trump} displays a fake tweet and Figure~\ref{fig:malicious_links_impersonation} shows an impersonation account's tweet with a suspicious link. Figure~\ref{fig:suggested_user} shows an example where \replaced[]{\emph{X}}{Twitter}'s suggestion algorithm returns as a first choice another username instead of the verified account.

\ignore{Figure~\ref{fig:David_Guetta_Impersonation} shows another impersonation attempt. }
\begin{figure}[H]
\centering
\includegraphics[width=\columnwidth,height=\textheight,keepaspectratio]{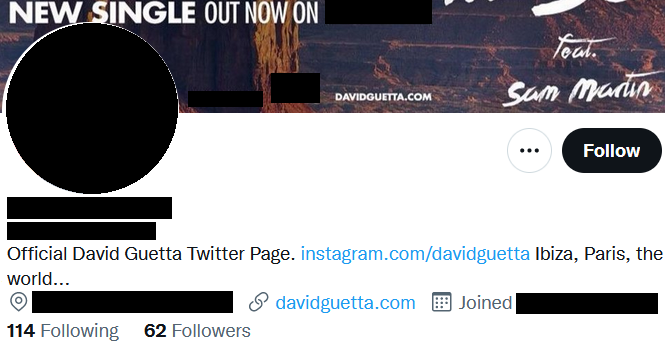}
\caption{An active squatted account which is a potential impersonation attempt of the artist David Guetta.}
\label{fig:David_Guetta_Impersonation}
\end{figure}

\begin{figure}[H]
\centering
\includegraphics[width=\columnwidth,height=\textheight,keepaspectratio]{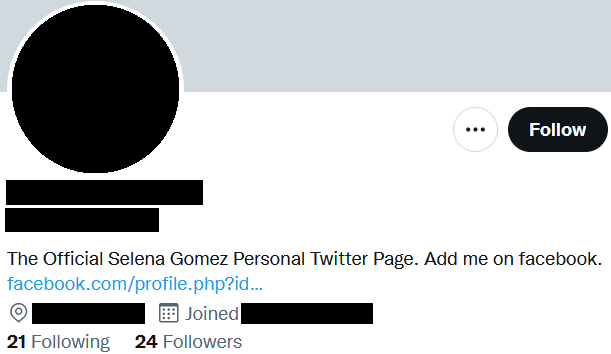}
\caption{The details of a squatted account attempting impersonation and posting suspicious links. }
\label{fig:malicious_links_profile}
\end{figure}

\begin{figure}[H]
\centering
\includegraphics[width=\columnwidth,height=\textheight,keepaspectratio]{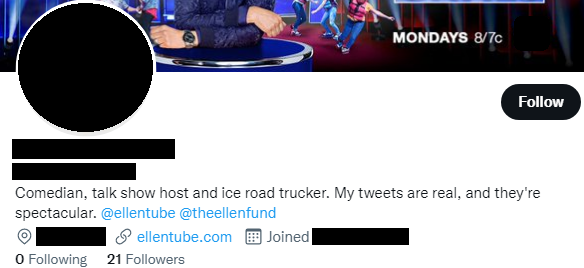}
\caption{An active squatted account which is a potential impersonation attempt of the comedian Ellen Lee DeGeneres.}
\label{fig:theellenshow_impersonation}
\end{figure}

\begin{figure}[H]
\centering
\includegraphics[width=\columnwidth,height=\textheight,keepaspectratio]{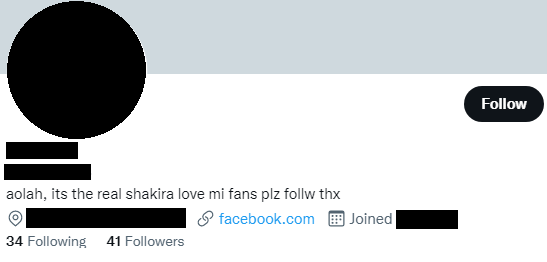}
\caption{An active squatted account which is a potential impersonation attempt of the artist Shakira.}
\label{fig:shakira_impersonation}
\end{figure}

Figures~\ref{fig:fake_trump} demonstrates the effect of a fake profile~\cite{Fake_Tweet}. 

\begin{figure}[H]
\centering
\includegraphics[width=\columnwidth,height=\textheight,keepaspectratio]{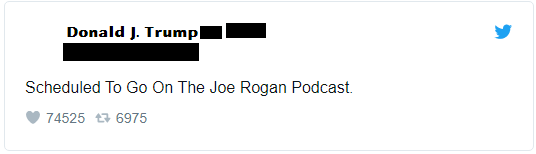}
\caption{A tweet of a fake profile with thousands of likes and retweets.}
\label{fig:fake_trump}
\end{figure}

\begin{figure}[H]
\centering
\includegraphics[width=\columnwidth,height=\textheight,keepaspectratio]{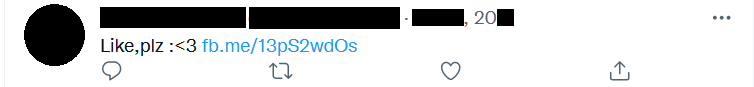}
\caption{An impersonation account tweeting a suspicious link. }
\label{fig:malicious_links_impersonation}
\end{figure}



\begin{figure}[H]
\centering
\includegraphics[width=\columnwidth,height=\textheight,keepaspectratio]{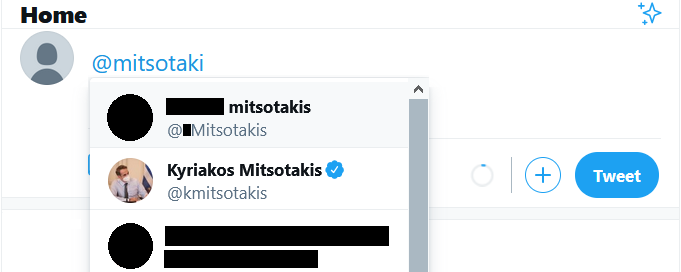}
\caption{An example where \replaced[]{\emph{X}}{Twitter}'s suggestion algorithm recommends as a first option a low popularity account instead of the verified one. We note that both accounts have the same edit distance from the mentioned word.}
\label{fig:suggested_user}
\vspace{-1em}
\end{figure}
}


\section{Image Similarity}
\label{appendix:image_similarity_algorithm}

In Section~\ref{subsec:amplification_of_confusion} we presented our image similarity experiment. Our image similarity algorithm is listed on Algorithm~\ref{alg:Image_Profile_Similarity}.


\begin{algorithm}[!htbp]
    \ignore{\small}
    \footnotesize
    \caption{Image Similarity Algorithm}\label{euclid}
    \begin{flushleft}
    \hspace*{\algorithmicindent} \textbf{Input} All usernames (original + generated) \\
    \hspace*{\algorithmicindent} \textbf{Output} Similar images
    \end{flushleft}
    \begin{algorithmic}[1]
        \Procedure{Image\_Similarity}{}
    
            \State $\textit{users[]} \gets \textit{original usernames / initial seed}$
            \If {$\textit{user not a celebrity}$}
                    \State Remove user from users
            \EndIf
        
            \For {user in range(len(users))}
        		\State $\textit{orig\_img[user]} \gets \textit{download image of celebrity}$
            \EndFor
        
            \For {user in range(len(orig\_img))}
        		\If {$\textit{image[user] does not have a face}$}
                    \State Remove orig\_img[user] from orig\_img[]
                    \State Remove user from users
                \EndIf
            \EndFor

            \State $\textit{gen\_users[][]} \gets \textit{generated username variants (UVs)}$ \\ \algorithmiccomment{1st array: all the remaining users} \\
            \algorithmiccomment{2nd: all the generated UVs of a user}
            \For {i in range(len(gen\_users))}
                \For{j in range(len(gen\_users[i]))}
                \State $\textit{gen\_imgs[i][j]} \gets \textit{download image of user}$ \\ \\ \algorithmiccomment{there are multiple images in each row}
        		\EndFor
            \EndFor
            
            \For {i in range(len(gen\_imgs))}
                \For {j in range(len(gen\_imgs[i]))} \\
                    \algorithmiccomment{calculation of distance between embeddings}
                    \If{distance$(orig\_img[i],gen\_imgs[i][j]) < $ threshold}
                        \State \Return Pair of similar images
                    \Else
                        \State No similarity
                    \EndIf
        		\EndFor
            \EndFor
        \EndProcedure
    \end{algorithmic}
    \label{alg:Image_Profile_Similarity}
\end{algorithm}

\ignore{

\section{Botometer Results}
\label{appendix:botometer_results}
In Section~\ref{subsec:bot_analysis} we examined whether a significant number of our squatted users are bots based on Botometer~\cite{Botometer}. Next, Figure~\ref{fig:bot_cap} demonstrates how likely the \replaced[]{\emph{X}}{Twitter} accounts are to be bots using the CAP score of Botometer\cite{Botometer}.


\begin{figure}[H]
\centering
\includegraphics[width=\columnwidth,height=\textheight,keepaspectratio]{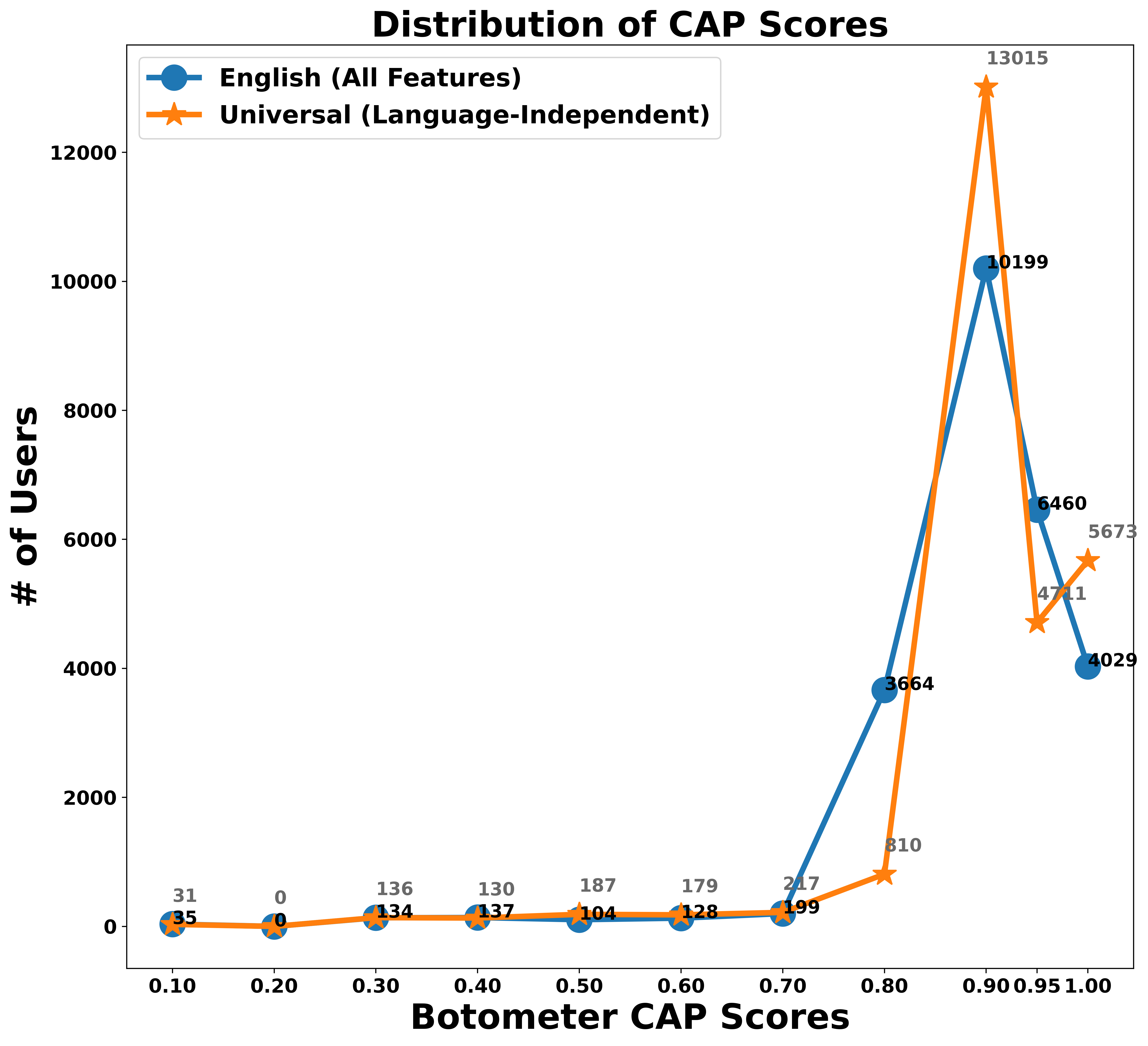}
\caption{Botometer: The distribution of \textit{CAP} scores across users.}
\label{fig:bot_cap}
\end{figure}

}

\ignore{
\begin{figure}[H]
\centering
\includegraphics[width=\columnwidth,height=\textheight,keepaspectratio]{images/bot_cap_line_plot.png}
\caption{Botometer: The distribution of \textit{CAP} scores across users.}
\label{fig:bot_cap}
\end{figure}
}

\ignore{

\section{Botometer CAP Score Selection}
\label{appendix:botometer_cap_selection}

In Section~\ref{subsec:bot_analysis} we applied Botometer to our squatted accounts by setting the \textit{CAP} score to $0.95$. Here, we provide more details on the selection of the specific \textit{CAP} score.

Setting the \textit{CAP} score to be $0.9$ ($10\%$ false positives) Botometer classified \textit{$20,873$} accounts as bots with minimum \textit{overall} raw score\footnote{Bot score in the $[0,1]$ range, using English (all features). In each case we have the overall score and the sub-scores for each bot class.} $0.94/1$. When setting the score to $0.95$ ($5\%$ false positives), $9,702$ profiles classified as bots with minimum \textit{overall} raw score $1$. To examine how conservative the thresholds are, we randomly selected $100$ accounts between the \textit{CAP} scores $0.9$ and $0.95$; out of those $68$ marked as ‘bots' and $32$ as ‘no bots'. Since more than $30\%$ of users marked as \textit{no bots} we discard the $0.9$ score. We then again randomly selected $100$ accounts with \textit{CAP} score from $0.95$ to $1$; out of those $89$ marked as ‘bots' and $11$ as ‘no bots'. We remark that our false positives cases are $11\%$ while Botometer claims $5\%$. For our analysis we chose a \textit{CAP} score of $0.95$ since it is a more conservative threshold than the $0.9$ and its false positives based on our sample are found to be acceptable.

}

\ignore{

\section{Botometer Analysis}
\label{appendix:botometer_analysis}

In Section~\ref{subsec:bot_analysis} we applied Botometer in order to classify our squatted accounts into different kinds of bots. Here, we discuss the most common classes. For each account, Botometer outputs different score values for every bot class, ranging from $0$ to $1$, which indicates the likeliness of it being a particular bot type. Figure~\ref{fig:bot_types} shows that a) most of the users belong to the ‘\textit{Fake follower}' and ‘\textit{Other}' categories, b) more than $5,000$ ‘English' profiles tend to be ‘\textit{Self-declared}' and ‘\textit{Spammer}' bots and c) the squatted accounts are unlikely to be political or financial bots. 

\begin{figure}[h!]
\centering
\includegraphics[width=\columnwidth,height=\textheight,keepaspectratio]{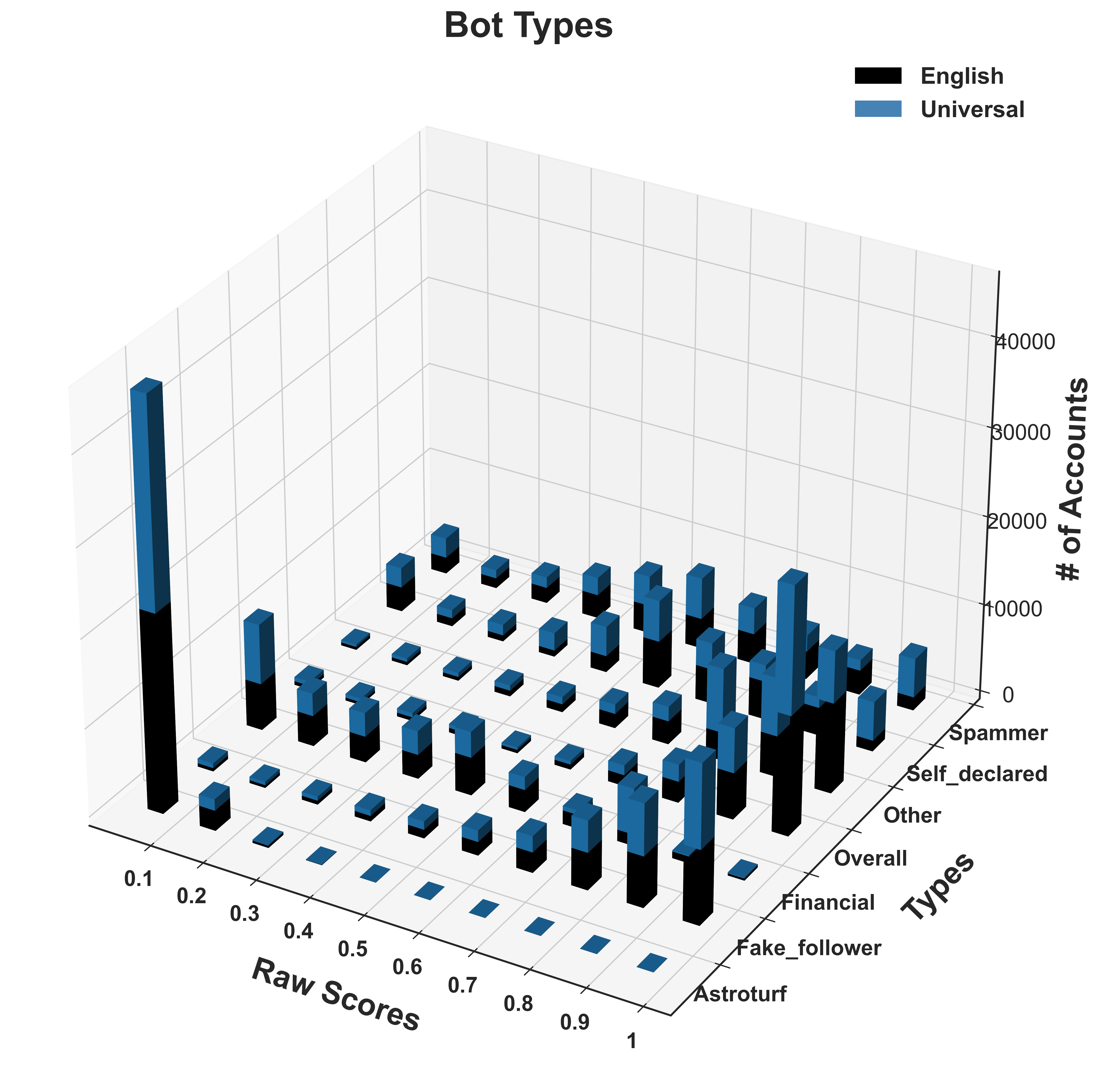}
\caption{Bot type categorization of squatted accounts.}
\label{fig:bot_types}
\end{figure}

}

\section{Profile Name Similarity}
\label{appendix:profile_name_similarity_algorithm}

In Section~\ref{subsec:amplification_of_confusion} we described our profile name similarity experiment. Our profile name similarity algorithm is described in Algorithm~\ref{alg:Profile_Name_Similarity}.

\begin{algorithm}[!htbp] 
    \footnotesize	
    
    \algblock[TryCatchFinally]{try}{endtry}
    \algcblock[TryCatchFinally]{TryCatchFinally}{finally}{endtry}
    \algcblockdefx[TryCatchFinally]{TryCatchFinally}{catch}{endtry}
        [1]{\textbf{catch} #1}
	{\textbf{end try}}
    \begin{flushleft}
        \hspace*{\algorithmicindent} \textbf{Input} All usernames (original + generated)\\
        \hspace*{\algorithmicindent} \textbf{Output} Similar profile names
    \end{flushleft}
    \begin{algorithmic}[1]
    \Procedure{Profile\_Name\_Similarity}{}
    \State $\textit{users[]} \gets \textit{original usernames / initial seed}$
    \State $\textit{gen\_users[][]} \gets \textit{generated UVs}$  \algorithmiccomment{1st array: all the users} \\
    \algorithmiccomment{2nd: all the generated UVs of a user}
    \For {i in range(len(gen\_users))}
        \try
            \\ \algorithmiccomment{For space limitations we assume that a profile name can be splitted by ‘ '}
            \State $\textit{prof\_name} \gets \textit{gen\_users[i][0].split(‘ ')}$
            \State $\textit{splitted\_name} \gets \textit{gen\_users[i][0].split(‘ ')}$
            \For{j in range(1, len(gen\_users[i]))}
            		\If {$prof\_name == \textit{gen\_users[i][j]}$}
                    \State \Return Exact match
                \ElsIf{$\textit{gen\_users[i][0]}$ in $\textit{gen\_users[i][j]}$ AND \\ \> \> \> \> \> \> \> \> \> \> \> \> \> \> $\textit{gen\_users[i][j]}$ not in $\textit{prof\_name}$}
                    \State \Return Exact match plus at least one \\
                    \> \> \> \> \> \> \> \> \> \> \> \> \> \> \> \> \> \> \> \> \> \> character extra \\
                \algorithmiccomment{For space limitations we assume that a profile name consists of 2 words}
                \ElsIf{$\textit{splitted\_name[0]}$ in $\textit{gen\_users[i][j]}$ AND \\ \> \> \> \> \> \> \> \> \> \> \> \> \> \> $\textit{splitted\_name[1]}$ not in  $\textit{gen\_users[i][j]}$} 
                    \State \Return At least one word of the $prof\_name$ \\
                    \> \> \> \> \> \> \> \> \> \> \> \> \> \> \> \> \> \> \> \> \> \> is appeared but in a different order
                \Else
                    \State \Return Other
                \EndIf
            \EndFor
        \catch{}
            \algorithmiccomment{We treat the profile name as one word only}
            \For{j in range(1, len(gen\_users[i]))}
            	\If {$prof\_name == \textit{gen\_users[i][j]}$}
                    \State \Return Exact match
                \ElsIf{\textit{gen\_users[i][0]} in \textit{gen\_users[i][j]} AND \\ \> \> \> \> \> \> \> \> \> \> \> \> \> \> \textit{gen\_users[i][j] not in prof\_name}}
                    \State \Return Exact match plus at least \\
                    \> \> \> \> \> \> \> \> \> \> \> \> \> \> \> \> \> \> \> \> \> \> one character extra
                \EndIf
            \EndFor
        \endtry
    \EndFor
    \EndProcedure
    \end{algorithmic}
    \caption{Profile Name Similarity Algorithm}
    \label{alg:Profile_Name_Similarity}
\end{algorithm}




\section{Non-Popular Accounts}
\label{appendix:non_popular_accounts}
In Section~\ref{subsec:squad_non_popular} we applied \frameworkName{} to a set of non-popular users. Here, Table~\ref{tab:non_popular_accounts} presents the classification results of \frameworkName{} on the $15$ non-popular accounts. \added[]{We observe that all the $61$ active accounts with a face in their profile photo were classified as \emph{benign} by \frameworkName{}. Lasty, \frameworkName{}{} returned only $1$ \emph{false negative} case.}

\begin{table}[!htbp]
\centering
\caption{\frameworkName{} results on non-popular accounts.}
\label{tab:non_popular_accounts}
\resizebox{\columnwidth}{!}{%
\begin{tabular}{cccccccccccc}
\hline
\multicolumn{12}{c}{\textbf{SQUAD - Non Popular Accounts}} \\ \hline
\textbf{\begin{tabular}[c]{@{}c@{}}Original\\ Account\end{tabular}} & \textbf{\begin{tabular}[c]{@{}c@{}}Generated \\ Accounts\end{tabular}} & \textbf{\begin{tabular}[c]{@{}c@{}}Accounts\\ with ED 1-3\end{tabular}} & \textbf{\begin{tabular}[c]{@{}c@{}}Active\\ Accounts\end{tabular}} & \textbf{\begin{tabular}[c]{@{}c@{}}Active\\ Accounts\\ with ED 1-3\end{tabular}} & \textbf{\begin{tabular}[c]{@{}c@{}}Suspended\\ Accounts\end{tabular}} & \textbf{\begin{tabular}[c]{@{}c@{}}Suspended\\ Accounts\\ with ED 1-3\end{tabular}} & \textbf{Followees} & \textbf{Followers} & \textbf{\begin{tabular}[c]{@{}c@{}}Classified\\ Malicious\\ by SQUAD\end{tabular}} & \textbf{\begin{tabular}[c]{@{}c@{}}Classified\\ Benign\\ by SQUAD\end{tabular}} & \textbf{\begin{tabular}[c]{@{}c@{}}Active Accounts\\  w/o Face in\\ Profile Image\end{tabular}} \\ \hline
Account1 & $11,876$ & $6,045$ ($50.9\%$) & $3$ ($0.02\%$) & $0$ ($0\%$) & $0$ ($0\%$) & $0$ ($0\%$) & $502$ & $186$ & $0$ ($0\%$) & $1$ ($33.3\%$) & $2$ ($66.6\%$) \\
Account2 & $5,975$ & $3,225$ ($53.9\%$) & $1$ ($0.01\%$) & $1$ ($100\%$) & $0$ ($0\%$) & $0$ ($0\%$) & $195$ & $15$ & $0$ ($0\%$) & $0$ ($0\%$) & $1$ ($100\%$) \\
Account3 & $12,929$ & $4,718$ ($36.4\%$) & $44$ ($0.34\%$) & $30$ ($68.1\%$) & $2$ ($0.01\%$) & $2$ ($100\%$) & $33$ & $94$ & $0$ ($0\%$) & $19$ ($43.2\%$) & $25$ ($56.8\%$) \\
Account4 & $6,983$ & $3,179$ ($45.5\%$) & $67$ ($0.96\%$) & $40$ ($59.7\%$) & $2$ ($0.03\%$) & $2$ ($100\%$) & $333$ & $69$ & $0$ ($0\%$) & $23$ ($34.3\%$) & $44$ ($65.7\%$) \\
Account5 & $11,612$ & $5,749$ ($49.5\%$) & $1$ ($0.01\%$) & $1$ ($100\%$) & $0$ ($0\%$) & $0$ ($0\%$) & $0$ & $0$ & $0$ ($0\%$) & $1$ ($100\%$) & $0$ ($0\%$) \\
Account6 & $1,002$ & $916$ ($91.4\%$) & $0$ ($0\%$) & $0$ ($0\%$) & $0$ ($0\%$) & $0$ ($0\%$) & $128$ & $145$ & $0$ ($0\%$) & $0$ ($0\%$) & $0$ ($0\%$) \\
Account7 & $4,541$ & $1,951$ ($42.9\%$) & $7$ ($0.15\%$) & $7$ ($100\%$) & $0$ ($0\%$) & $0$ ($0\%$) & $4$ & $4$ & $0$ ($0\%$) & $3$ ($42.9\%$) & $4$ ($57.1\%$) \\
Account8 & $1,454$ & $1,287$ ($88.5\%$) & $0$ ($0\%$) & $0$ ($0\%$) & $0$ ($0\%$) & $0$ ($0\%$) & $173$ & $65$ & $0$ ($0\%$) & $0$ ($0\%$) & $0$ ($0\%$) \\
Account9 & $13,756$ & $6,883$ ($50\%$) & $23$ ($0.16\%$) & $18$ ($78.3\%$) & $1$ ($0.01\%$) & $1$ ($100\%$) & $0$ & $0$ & \begin{tabular}[c]{@{}c@{}}$0$ ($0\%$)\\ $\#FP = 0$\end{tabular} & {\color[HTML]{333333} \begin{tabular}[c]{@{}c@{}}$6$ ($26.1\%$)\\ $\#FN = 1$\end{tabular}} & $17$ ($73.9$\%) \\
Account10 & $3,116$ & $2,860$ ($91.8\%$) & $0$ ($0\%$) & $0$ ($0\%$) & $0$ ($0\%$) & $0$ ($0\%$) & $121$ & $129$ & $0$ ($0\%$) & $0$ ($0\%$) & $0$ ($0\%$) \\
Account11 & $4,159$ & $2,854$ ($68.6\%$) & $3$ ($0.07$) & $3$ ($100\%$) & $0$ ($0\%$) & $0$ ($0\%$) & $79$ & $140$ & $0$ ($0\%$) & $1$ ($33.3\%$) & $2$ ($66.6\%$) \\
Account12 & $1,779$ & $1,644$ ($92.4\%$) & $2$ ($0.11\%$) & $2$ ($100\%$) & $0$ ($0\%$) & $0$ ($0\%$) & $14$ & $6$ & $0$ ($0\%$) & $2$ ($100\%$) & $0$ ($0\%$) \\
Account13 & $8,164$ & $3,885$ ($47.6\%$) & $7$ ($0.09\%$) & $5$ ($71.4\%$) & $0$ ($0\%$) & $0$ ($0\%$) & $490$ & $119$ & $0$ ($0\%$) & $4$ ($57.1\%$) & $3$ ($42.9\%$) \\
Account14 & $6,360$ & $4,627$ ($72.8\%$) & $0$ ($0\%$) & $0$ ($0\%$) & $0$ ($0\%$) & $0$ ($0\%$) & $78$ & $28$ & $0$ ($0\%$) & $0$ ($0\%$) & $0$ ($0\%$) \\
Account15 & $7,883$ & $4,524$ ($57.4\%$) & $1$ ($0.01\%$) & $0$ ($0\%$) & $0$ ($0\%$) & $0$ ($0\%$) & $62$ & $38$ & $0$ ($0\%$) & $1$ ($100\%$) & $0$ ($0\%$) \\ \hline
\textbf{Total} & \textbf{$101,589$} & \textbf{\begin{tabular}[c]{@{}c@{}}$54,347$ \\ ($53.5\%$)\end{tabular}} & \textbf{\begin{tabular}[c]{@{}c@{}}$159$ \\ ($15.7\%$)\end{tabular}} & \textbf{\begin{tabular}[c]{@{}c@{}}$107$ \\ ($67.3\%$)\end{tabular}} & \textbf{\begin{tabular}[c]{@{}c@{}}$5$ \\ ($4.92\%$)\end{tabular}} & \textbf{\begin{tabular}[c]{@{}c@{}}$5$ \\ ($100\%$)\end{tabular}} & \textbf{$2,212$} & \textbf{$1,038$} & \textbf{\begin{tabular}[c]{@{}c@{}}$0$ ($0\%$)\\ $\#FP = 0$\end{tabular}} & \textbf{\begin{tabular}[c]{@{}c@{}}$61$ ($38.4\%$)\\ $\#FN = 1$\end{tabular}} & \textbf{\begin{tabular}[c]{@{}c@{}}$98$ \\ ($61.6\%$)\end{tabular}} \\ \hline
\end{tabular}}
\end{table}

\ignore{
\tl{\section{Popular Accounts}}
\label{appendix:popular_accounts}
\tl{Table~\ref{tab:popular_accounts} presents the classification results of \frameworkName{} on $64$~\footref{ref:64_accounts} popular accounts of our \emph{Initial Seed}. The profiles are grouped according Twitter's accepted account types~\cite{Twitter_Badge} (see Section~\ref{subsec:meth_mq1}).} 
}

\section{Username Characteristics}
\label{appendix:username_characteristics}

\begin{figure}[!htbp]
    \centering
    \includegraphics[width=\columnwidth,height=\textheight,keepaspectratio]{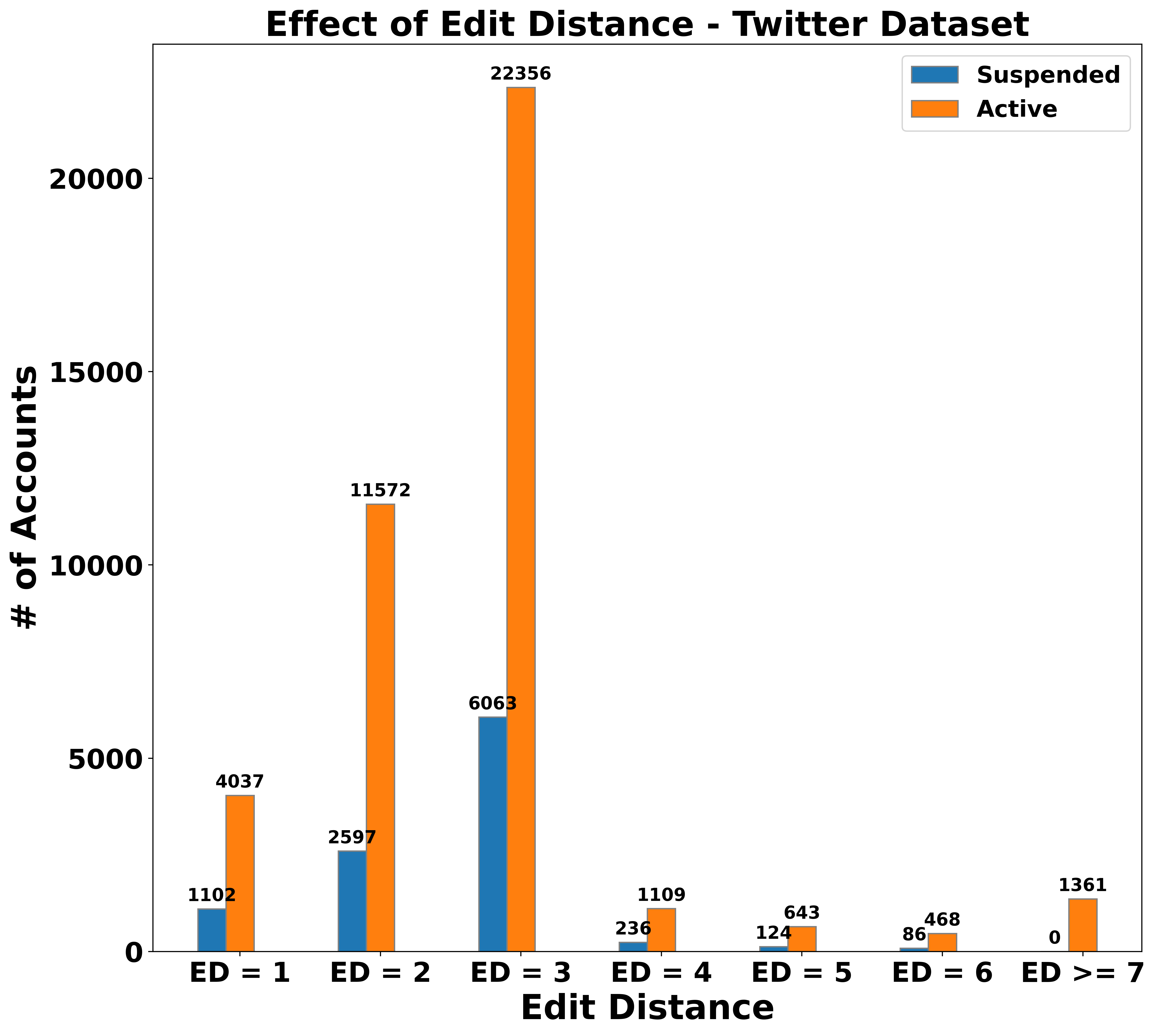}
    \caption{The number of active and suspended accounts across different edit distance categories. }
    \label{fig:Twitter_suspended_existing_edit_distance}
\end{figure}

In Section~\ref{subsec:characteristics} we analyzed the characteristics of our squatted usernames. Figure~\ref{fig:Twitter_suspended_existing_edit_distance} further depicts the number of suspended and active accounts over different edit distance values. \added[]{We observe that the highest number of both active and suspended profiles are squatted usernames with edit distance one to three characters from their target.}

\section{\frameworkName{}'s Performance}
\label{appendix:squad_performance}

In Section~\ref{subsec:classifier_performance} we discussed the training and evaluation metrics applied for \frameworkName{} as well as the overall performance of all the classifiers. Here we further present the actual results of the classification performance of all the algorithms. Table~\ref{tab:classification_report} depicts the classification performance results of the $7$ models. Random Forest exhibits the best performance in identifying suspicious accounts.

\begin{table}[!htbp]
\caption{Classification performance. (P) shows the positive class.}
\label{tab:classification_report}
\centering
\resizebox{\columnwidth}{!}{%
\begin{tabular}{cccc}
\hline
\textbf{Models} &
  \multicolumn{1}{c}{\textbf{Precision}} &
  \multicolumn{1}{c}{\textbf{Recall}} &
  \textbf{F1 - Score} \\ \hline
\textbf{Random Forest} &
  \multicolumn{3}{l}{} \\ 
\multicolumn{1}{r}{\begin{tabular}[c]{@{}r@{}}(P) Suspicious\\ (P) Benign\end{tabular}} &
  \multicolumn{1}{c}{\begin{tabular}[c]{@{}c@{}}$0.97$\\ $0.91$\end{tabular}} &
  \multicolumn{1}{c}{\begin{tabular}[c]{@{}c@{}}$0.94$\\ $0.95$\end{tabular}} &
  \begin{tabular}[c]{@{}c@{}}$0.95$\\ $0.93$\end{tabular} \\ 
\textbf{Naive Bayes} &
  \multicolumn{3}{l}{} \\ 
\multicolumn{1}{r}{\begin{tabular}[c]{@{}r@{}}(P) Suspicious\\ (P) Benign\end{tabular}} &
  \multicolumn{1}{c}{\begin{tabular}[c]{@{}c@{}}$0.98$\\ $0.78$\end{tabular}} &
  \multicolumn{1}{c}{\begin{tabular}[c]{@{}c@{}}$0.82$\\ $0.97$\end{tabular}} &
  \begin{tabular}[c]{@{}c@{}}$0.89$\\ $0.86$\end{tabular} \\ 
\textbf{Logistic Regression} &
  \multicolumn{3}{l}{} \\ 
\multicolumn{1}{r}{\begin{tabular}[c]{@{}r@{}}(P) Suspicious\\ (P) Benign\end{tabular}} &
  \multicolumn{1}{c}{\begin{tabular}[c]{@{}c@{}}$0.94$\\ $0.85$\end{tabular}} &
  \multicolumn{1}{c}{\begin{tabular}[c]{@{}c@{}}$0.90$\\ $0.91$\end{tabular}} &
  \begin{tabular}[c]{@{}c@{}}$0.92$\\ $0.88$\end{tabular} \\ 
\textbf{SVM} &
  \multicolumn{3}{l}{} \\ 
\multicolumn{1}{r}{\begin{tabular}[c]{@{}r@{}}(P) Suspicious\\ (P) Benign\end{tabular}} &
  \multicolumn{1}{c}{\begin{tabular}[c]{@{}c@{}}$0.95$\\ $0.85$\end{tabular}} &
  \multicolumn{1}{c}{\begin{tabular}[c]{@{}c@{}}$0.90$\\ $0.93$\end{tabular}} &
  \begin{tabular}[c]{@{}c@{}}$0.92$\\ $0.89$\end{tabular} \\ 
\textbf{Decision Trees} &
  \multicolumn{3}{l}{} \\ 
\multicolumn{1}{r}{\begin{tabular}[c]{@{}r@{}}(P) Suspicious\\ (P) Benign\end{tabular}} &
  \multicolumn{1}{c}{\begin{tabular}[c]{@{}c@{}}$0.94$\\ $0.90$\end{tabular}} &
  \multicolumn{1}{c}{\begin{tabular}[c]{@{}c@{}}$0.93$\\ $0.91$\end{tabular}} &
  \begin{tabular}[c]{@{}c@{}}$0.94$\\ $0.90$\end{tabular} \\ 
\textbf{K-Nearest Neighbor} &
  \multicolumn{3}{l}{} \\ 
\multicolumn{1}{r}{\begin{tabular}[c]{@{}r@{}}(P) Suspicious\\ (P) Benign\end{tabular}} &
  \multicolumn{1}{c}{\begin{tabular}[c]{@{}c@{}}$0.93$\\ $0.81$\end{tabular}} &
  \multicolumn{1}{c}{\begin{tabular}[c]{@{}c@{}}$0.86$\\ $0.90$\end{tabular}} &
  \begin{tabular}[c]{@{}c@{}}$0.89$\\ $0.85$\end{tabular} \\ 
\textbf{Neural Network} &
  \multicolumn{3}{l}{} \\ 
\multicolumn{1}{r}{\begin{tabular}[c]{@{}r@{}}(P) Suspicious\\ (P) Benign\end{tabular}} &
  \multicolumn{1}{c}{\begin{tabular}[c]{@{}c@{}}$0.96$\\ $0.86$\end{tabular}} &
  \multicolumn{1}{c}{\begin{tabular}[c]{@{}c@{}}$0.90$\\ $0.94$\end{tabular}} &
  \begin{tabular}[c]{@{}c@{}}$0.93$\\ $0.90$\end{tabular} \\ \hline
\end{tabular}
}
\end{table}

\begin{figure}[!htbp]
\centering
\includegraphics[width=\columnwidth,height=\textheight,keepaspectratio]{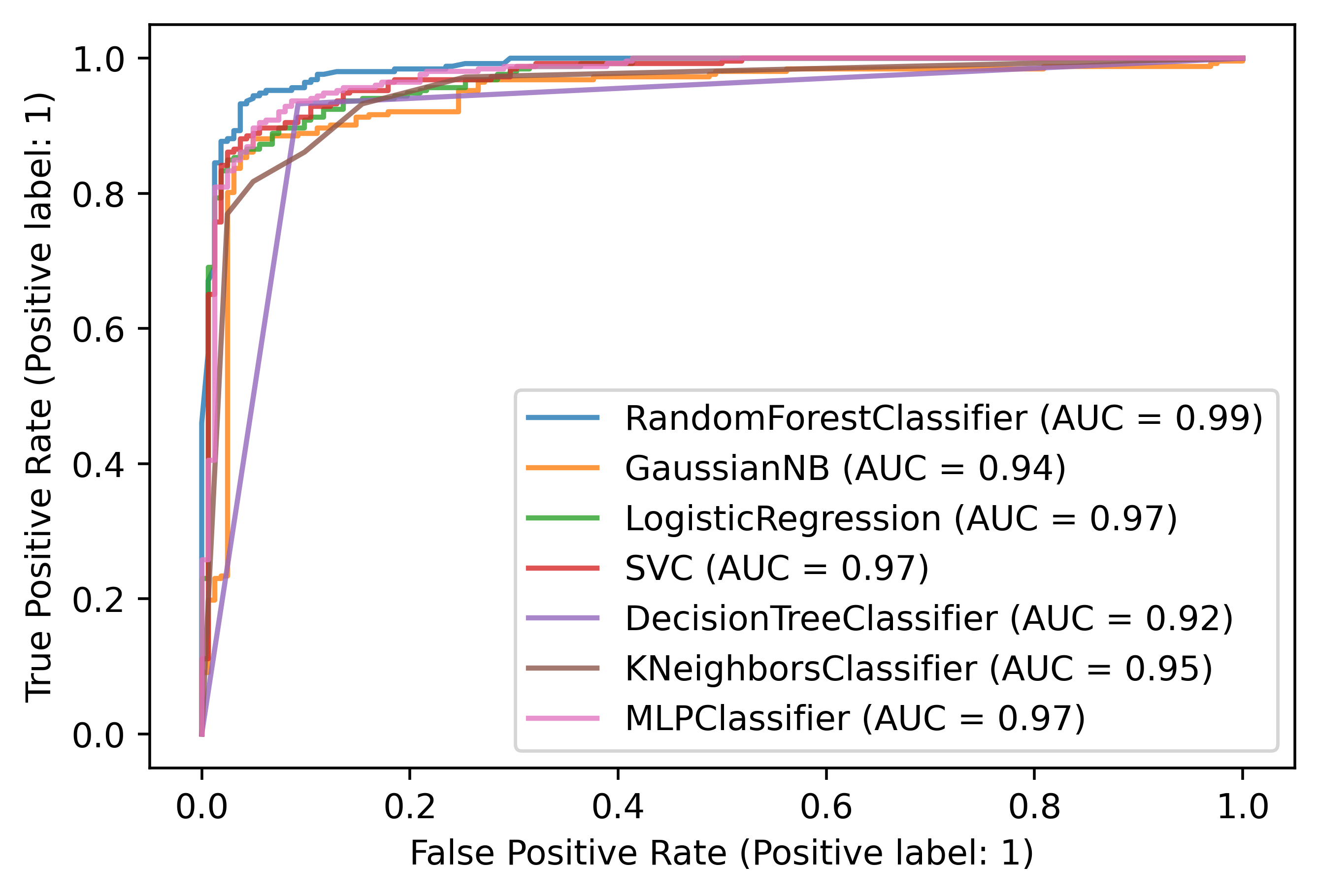}
\caption{ROC Curve of all the classification models.}
\label{fig:roc_curve}
\end{figure}

Figure~\ref{fig:roc_curve} shows the ‘\textit{Mean AUC}' score of all classification models.

\ignore{

\section{Username Squatting on other Platforms}
\label{appendix:username_squatting_on_other_platforms}

In Section~\ref{subsec:username_squatting_across_platforms} we discussed ways to identify if username squatting is prevalent in other social platforms. Next, we provide more details about our measurement methodology and findings. We observe that a lot of popular accounts maintain accounts on other platforms with the same username. Therefore, if other platforms are plagued with similar issues, then it is likely that variants that were marked as malicious on \replaced[]{\emph{X}}{Twitter} could appear on the other platforms too. To examine this, we select the top $10$ accounts of our seed and \added[]{manually} check how many exist with the same username across both Instagram and TikTok, remaining with $5$ profiles. Next, we examine whether the username variants\added[]{, previously generated by \frameworkName{},} \replaced[]{for}{of those} profiles\deleted[]{that are} manually identified as impersonators on \replaced[]{\emph{X}}{Twitter} a) exist with the same username, b) have a similar image and c) perform potential malicious activity on other OSNs. We call ‘\emph{Level $\textit{1}$}' the users who cover only the first property, ‘\emph{Level $\textit{2}$}' the profiles which cover the first property plus any of the other two and ‘\emph{Level $\textit{3}$}' the accounts that cover all the aforementioned properties. 

Table~\ref{tab:impersonation_across_OSNs} summarises our findings.
We find that the most common pattern of the active squatted accounts across all OSNs remains the \emph{Number Insertion} pattern. Lastly\ignore{More importantly}, we observe that almost half of the squatted accounts that impersonate on \replaced[]{\emph{X}}{Twitter} exist on other OSNs and have common properties. 


\begin{table}[H]
\caption{Similar impersonation attempts across different Social Networks.}
\label{tab:impersonation_across_OSNs}
\resizebox{\columnwidth}{!}{%
\begin{tabular}{lcccc}
\hline
\multicolumn{5}{c}{\textbf{Impersonations Across OSNs}} \\ \hline
\multicolumn{1}{c}{\textbf{Users}} &
  \multicolumn{1}{c}{\textbf{\begin{tabular}[c]{@{}c@{}}\replaced[]{\emph{X}}{Twitter} - Active\\ Impersonations\end{tabular}}} &
  \multicolumn{1}{c}{\textbf{Category}} &
  \multicolumn{1}{c}{\textbf{Instagram}} &
  \textbf{TikTok} \\ \hline
\multicolumn{1}{l}{\multirow{2}{*}{@katyperry}} &
  \multicolumn{1}{c}{\multirow{2}{*}{$7$}} &
  \multicolumn{1}{c}{Level $3$} &
  \multicolumn{1}{c}{$1/7$} &
  $0/7$ \\ 
\multicolumn{1}{l}{} &
  \multicolumn{1}{c}{} &
  \multicolumn{1}{c}{Level $2$} &
  \multicolumn{1}{c}{$4/7$} &
  $4/7$ \\ \cline{3-5} 
\multicolumn{1}{l}{\multirow{2}{*}{@justinbieber}} &
  \multicolumn{1}{c}{\multirow{2}{*}{$20$}} &
  \multicolumn{1}{c}{Level $3$} &
  \multicolumn{1}{c}{$3/20$} &
  $4/20$ \\ 
\multicolumn{1}{l}{} &
  \multicolumn{1}{c}{} &
  \multicolumn{1}{c}{Level $2$} &
  \multicolumn{1}{c}{$7/20$} &
  $10/20$ \\ \cline{3-5} 
\multicolumn{1}{l}{\multirow{2}{*}{@cristiano}} &
  \multicolumn{1}{c}{\multirow{2}{*}{$1$}} &
  \multicolumn{1}{c}{Level $3$} &
  \multicolumn{1}{c}{$0/1$} &
  $0/1$ \\ 
\multicolumn{1}{l}{} &
  \multicolumn{1}{c}{} &
  \multicolumn{1}{c}{Level $2$} &
  \multicolumn{1}{c}{$1/1$} &
  $1/1$ \\\cline{3-5} 
\multicolumn{1}{l}{\multirow{2}{*}{@ladygaga}} &
  \multicolumn{1}{c}{\multirow{2}{*}{$10$}} &
  \multicolumn{1}{c}{Level $3$} &
  \multicolumn{1}{c}{$0/10$} &
  $1/10$ \\ 
\multicolumn{1}{l}{} &
  \multicolumn{1}{c}{} &
  \multicolumn{1}{c}{Level $2$} &
  \multicolumn{1}{c}{$3/10$} &
  $7/10$ \\ \cline{3-5} 
\multicolumn{1}{l}{\multirow{2}{*}{@kimkardashian}} &
  \multicolumn{1}{c}{\multirow{2}{*}{$5$}} &
  \multicolumn{1}{c}{Level $3$} &
  \multicolumn{1}{c}{$2/5$} &
  $0/5$ \\ 
\multicolumn{1}{l}{} &
  \multicolumn{1}{c}{} &
  \multicolumn{1}{c}{Level $2$} &
  \multicolumn{1}{c}{$3/5$} &
  $0/5$ \\ \cline{3-5} 
\multicolumn{1}{c}{\multirow{3}{*}{\textbf{Total}}} &
  \multicolumn{1}{c}{\multirow{3}{*}{\textbf{$\textbf{42}$}}} &
  \multicolumn{1}{c}{\textbf{Level $\textbf{3}$}} &
  \multicolumn{1}{c}{\textbf{$\textbf{6/42}$}} &
  \textbf{$\textbf{5/42}$} \\ 
\multicolumn{1}{c}{} &
  \multicolumn{1}{c}{} &
  \multicolumn{1}{c}{\textbf{Level $\textbf{2}$}} &
  \multicolumn{1}{c}{\textbf{$\textbf{18/42}$}} &
  \textbf{$\textbf{22/42}$} \\ 
\multicolumn{1}{c}{} &
  \multicolumn{1}{c}{} &
  \multicolumn{1}{c}{\textbf{Level $\textbf{1}$}} &
  \multicolumn{1}{c}{\textbf{$\textbf{27/42}$}} &
  \textbf{$\textbf{28/42}$} \\ \hline
\end{tabular}}
\end{table}

}

\section{Ethics: Responsible Disclosure}
\label{appendix:ethics}
\ignore{\vspace{3pt}\noindent\textbf{Dataset Collection.}
In this work we analyzed data from the \replaced[]{\emph{X}}{Twitter} network and demonstrated that username squatting can contribute to online confusion. We did not extract any personal opinions or other viewpoints linked to individuals as this could potentially be sensitive. In detail, a) all our collected raw data (profile data, tweets with mentions etc.) were captured using official APIs and b) all the unnecessary data that APIs return are discarded immediately. We also b) never share any information with $3$rd parties and c) encrypt and keep in a local disk all the data after each experiment. We furthermore follow recommendations in~\cite{williams2017towards} and mask the account names of individuals in the paper and do not include any tweets of individual accounts, including only public tweets from the top $97$ most popular users.}
\replaced[]{}{
\vspace{3pt}\noindent\textbf{Responsible Disclosure.}}
We believe it should be the responsibility of the platform to integrate automated processes to identify impersonators and suspicious accounts. We reported all the manually verified \emph{impersonators} (see Section~\ref{subsec:bot_analysis}) to \replaced[]{\emph{X}}{Twitter} via a formal report at HackerOne program~\cite{HackerOne}. \replaced[]{\emph{X}}{Twitter} acknowledged the problem and marked our report as \emph{Informative}. Unfortunately, \replaced[]{\emph{X}}{Twitter}'s response (see below) suggests that the platform simply relies on weak crowdsourcing methods~\cite{Twitter_response_impersonation} for identifying impersonation attempts even though it is aware of the problem. Such methods are known to suffer from erroneous reports and manual verification of crowdsourced reports does not scale.

We share below \replaced[]{\emph{X}}{the Twitter}'s official response (verbatim) to our report which included all manually verified and active impersonation accounts:
``Thank you for your report.
Please bear in mind that our HackerOne program~\cite{HackerOne} is for the reporting of explicit security vulnerabilities in Twitter services. We are already aware that accounts impersonating high-profile Twitter users is an issue on our platform. For this reason, we already provide additional controls~\cite{Twitter_response_impersonation} which allow users to report these accounts which may be in violation of Twitter rules~\cite{Twitter_response_rules}.
While this is an issue we are already aware of, this report does not appear to concern an exploitable vulnerability in Twitter services. Furthermore, as we already have controls in place to allow Twitter users to report impersonator accounts, we will be closing this report as ‘Informative'.
Regardless, we do appreciate your interest in our program, and we encourage you to continue hunting for security vulnerabilities.
Thank you for thinking of Twitter security.''
